\documentclass[fleqn,usenatbib]{mnras}

\usepackage{newtxtext,newtxmath}

\usepackage[T1]{fontenc}

\DeclareRobustCommand{\VAN}[3]{#2}
\let\VANthebibliography\thebibliography
\def\thebibliography{\DeclareRobustCommand{\VAN}[3]{##3}\VANthebibliography}

\usepackage{mathrsfs}

\usepackage{graphicx}	
\usepackage{amsmath}	
\usepackage{bm}

\newcommand{\HI}{H{\sc~i}}

\newcommand{\pd}[2]{\frac{\partial #1}{\partial #2}}

\newcommand{\unifC}{{\UrlFont Uniform $\text{\UrlFont C}_{\text{\UrlFont R}}$}}
\newcommand{\maxC}{{\UrlFont Maximum $\text{\UrlFont C}_{\text{\UrlFont R}}(\Delta)$}}
\newcommand{\unifCb}{{\UrlFont Uniform $\text{\bf \UrlFont C}_{\text{\bf \UrlFont R}}$}~}
\newcommand{\maxCb}{{\UrlFont Maximum $\text{\bf \UrlFont C}_{\text{\bf \UrlFont R}}({\bm \Delta})$}}

\title[Morphology of Reionization]{The Morphology of Reionization in a Dynamically Clumpy Universe}

\graphicspath{{./}{figures/}}

\author[Cain et al.]{
Christopher Cain,$^{1}$\thanks{E-mail: ccain002@ucr.edu}
Anson D'Aloisio,$^{1}$
Nakul Gangolli$^{1}$
and Matthew McQuinn$^{2}$
\\
$^{1}$Department of Physics and Astronomy, University of California, Riverside, CA 92521, USA\\
$^{2}$Department of Astronomy, University of Washington, Seattle, WA 98195-1580, USA
}

\date{Accepted XXX. Received YYY; in original form ZZZ}

\pubyear{2022}

\begin{document}
\label{firstpage}
\pagerange{\pageref{firstpage}--\pageref{lastpage}}
\maketitle

\begin{abstract}  
A recent measurement of the Lyman-limit mean free path at $z = 6$ suggests it may have been very short, motivating a better understanding of the role that ionizing photon sinks played in reionization.  Accurately modeling the sinks in reionization simulations is challenging because of the large dynamic range required if $\sim 10^4-10^8 M_{\odot}$ gas structures  contributed significant opacity. Thus, there is no consensus on how important the sinks were in shaping reionization's morphology.    We address this question with a recently developed radiative transfer code that includes a dynamical sub-grid model for the sinks based on radiative hydrodynamics simulations.  Compared to assuming a fully pressure-smoothed IGM, our dynamical treatment reduces ionized bubble sizes by $10-20\%$ under typical assumptions about reionization's sources.  Near reionization’s midpoint, the 21 cm power at $k \sim 0.1$ $h$Mpc$^{-1}$ is similarly reduced.  These effects are more modest than the $30-60\%$ suppression resulting from the higher recombination rate if pressure smoothing is neglected entirely.  Whether the sinks played a significant role in reionization's morphology depends on the nature of its sources. For example, if reionization was driven by bright ($M_{\rm UV} < -17$) galaxies, the sinks reduce the large-scale 21 cm power by at most $20\%$, even if pressure smoothing is neglected. Conveniently, when bright sources contribute significantly, the morphology in our dynamical treatment can be reproduced accurately with a uniform sub-grid clumping factor that yields the same ionizing photon budget.  By contrast, if $M_{\rm UV} \sim -13$ galaxies drove reionization, the uniform clumping model can err by up to $40\%$. 

\end{abstract}




\section{Introduction} \label{sec:intro}

The past decade has seen an increase in the number and quality of observational constraints on the Epoch of Reionization (EoR).  Planck's measurement of the cosmic microwave background (CMB) Thomson scattering optical depth ($\tau_{\rm es}$) have revised the midpoint of reionization to $z \approx 7.5$, driving the field toward late reionization models~\citep{Planck2018}.  Meanwhile, studies of damping wings in high-z quasar spectra~\citep{Mortlock2011, Grieg2016, Davies2018} and Lyman Alpha Emitter (LAE) surveys~\citep{Kashikawa2006, Ono2011, Schenker2012, Pentericci2014, Mesigner2015, Ouchi2018, Hu2019} have also suggested a significantly neutral intergalactic medium (IGM) at $z\sim 7$.  At $z\lesssim 6$, quasar absorption spectra measurements may also be consistent with an ongoing reionization process down to $z\sim 5$ \citep[e.g.][]{Becker2015,Kulkarni2019,2021MNRAS.506.2390Q, Bosman2021,Zhu2021}.
Future observations with the James Webb Space Telescope (JWST), the extremely large telescopes, 21 cm signal experiments -- e.g. SKA \citep{Mellema2013} and HERA~\citep{HERA2021b, HERA2021a} -- and other line intensity mapping surveys (e.g. SPHEREx; \citealt{Dore2014}), promise to vastly expand our understanding of the EoR.  This wealth of forthcoming data motivates theoretical studies to predict and interpret reionization observables with greater accuracy.  

All reionization observables, with the exception of $\tau_{\rm es}$, are sensitive to the spatial structure of ionized regions, broadly termed morphology.  Reionization's morphology is known to be sensitive to the nature of its sources as well as the LyC opacity of the IGM \citep{Furlanetto2005,Iliev2005b,McQuinn2007,Alvarez2012,Sobacchi2014,Davies2021}. During reionization, gaseous halos with masses $\lesssim 10^8M_{\odot}$, which are too small to form stars, act as sinks of ionizing photons and play a role in setting the IGM opacity~\citep{Shapiro2004,Iliev2005b}.  The sinks can be as small as $10^4 M_{\odot}$ before reionization, roughly the Jeans filtering scale in the cold IGM~\citep{Gnedin2000,Naoz2007,Emberson2013}.  Once the IGM surrounding these structures ionizes, their gas is photo-evaporated and pressure-smoothed over a timescale of a few hundred Myr~\citep{Iliev2005,Park2016,DAloisio2020,Nasir2021}.  We refer to this process as relaxation. Modeling relaxation in simulations requires high ($\sim$ kpc) spatial resolution to resolve the sinks~\citep{Emberson2013} and  radiative transfer (RT) coupled to the hydrodynamics to capture the interplay between self-shielding and pressure smoothing~\citep{Park2016,DAloisio2020}.

In RT simulations that are big enough to capture the large-scale structure of patchy reionization~\citep[$\gtrsim$ 200-300 Mpc,][]{Iliev2014,Kaur2020}, resolving the sinks presents an extreme computational challenging owing to the $> 5$ orders of magnitude in spatial scales that are required.  RT simulations that come close~\citep[e.g.][]{Gnedin2014,Ocvirk2016,Kannan2022} are too expensive to run more than a handful of times. On the other hand, the semi-numerical methods of approximating RT that have been employed for parameter space studies either ignore the effect of the sinks or model them in an approximate manner \citep[e.g.][]{Choudhury2021,Gazagnes2021,Davies2021}.  
It is unclear, however, which approximation schemes for the sinks are accurate.  Simulations that ignore the unresolved sinks implicitly assume that their effects are fully degenerate with the parameters that characterize the sources \citep[][]{Iliev2005b}. Other studies have attempted to model unresolved sinks with a sub-grid clumping factor~\citep[][]{McQuinn2007,Mao2019}, by adding extra opacity to their cells~\citep[][]{Shukla2016,Giri2019}, or by specifying the mean free path as an input \citep[][]{Davies2016,Wu2021b,Davies2021,Trac2021}.  These implementations vary in complexity and often disagree on what role the sinks play. As a result, currently there is no consensus on how much of an effect the sinks have on reionization and, relatedly, how important they are for interpreting observables.  This paper aims to further address these questions. 

Another motivation for the current study is the recent measurement of the Lyman-Limit mean free path at $z = 6$ by \citealt{Becker2021} (see also \citealt{Bosman2021b} for complementary constraints).  They reported a value of $\lambda_{912}^{\rm mfp} = 3.57^{+3.09}_{-2.14}$ $h^{-1}$cMpc, which is considerably shorter than extrapolations from measurements at lower redshift \citep{Worseck2014}.  In addition to suggesting that the IGM may have still been significantly neutral at $z = 6$ \citep{Cain2021,Garaldi2022,Lewis2022}, their measurement -- if confirmed -- may indicate that absorptions in ionized gas consumed a majority of the reionization photon budget~\citep{Davies2021b}; in which case, accounting for the effect of sinks in simulations would be critical.  

The main goal of this work is to assess how important the sinks are for modeling reionization's morphology. 
Towards this end, we use a new ray-tracing RT code that was first applied in \citet{Cain2021}.  The code has been developed for flexibility and low computational cost, mainly by the use of large cell sizes and adjustable angular resolution in the RT calculation.  For our fiducial simulations, we employ the \citet{Cain2021} sub-grid model based on a suite of high-resolution, fully coupled hydro/RT simulations, which track how the LyC opacity of the IGM evolves in different environments after I-fronts sweep through (an expanded version of the numerical experiments in \citealt{DAloisio2020}).  However, one of the main features of our RT code is that any sub-grid model of IGM opacity can be straightforwardly implemented.  We exploit this feature to compare the reionization morphologies in our detailed fiducial simulations against sink models constructed to mimic the various assumptions made previously in the literature. 

Another goal of this work is to explore the relationship between reionization sources and sinks.  The large uncertainty in the nature of the LyC sources necessitates exploring the sinks in different source models. Although it is widely believed that galaxies were the main drivers of reionization, it remains unclear which galaxies sourced the LyC background~\citep[see for example][]{Robertson2015,Finkelstein2019,Naidu2020,Lewis2020}. A number of studies have looked at the impact of different models for the sources and sinks separately; to our knowledge none have directly addressed the interplay between the two.

This work is organized as follows. In \S\ref{sec:methods}, we describe our numerical methods. In \S\ref{sec:results} we study the morphology of reionization in different sinks models.  In \S\ref{sec:sources}, we extend our analysis to include different models for the sources.  We summarize our results and conclude in \S\ref{sec:conc}.  Throughout this work, we assume the following cosmological parameters: $\Omega_m = 0.305$, $\Omega_{\Lambda} = 1 - \Omega_m$, $\Omega_b = 0.048$, $h = 0.68$, $n_s = 0.9667$ and $\sigma_8 = 0.82$, consistent with the~\citet{Planck2018} results. All distances are quoted in comoving units unless otherwise specified.  

\section{Numerical Methods} \label{sec:methods}
 
\subsection{Large-Scale Radiative Transfer} \label{subsec:RT}

We ran our reionization simulations using the new RT code of~\citet{Cain2021}.  Here we describe the features of the code relevant for this work, leaving a more detailed presentation to a future paper.  

The code inputs are a time-series of halo catalogs and coarse-grained density fields from a cosmological N-body simulation.  Halos are assigned ionizing photon emissivities and binned to their nearest grid points on the RT grid.  Rays are cast from the centers of source cells at each time step.  As rays travel, the optical depth through each cell is computed and photons are deposited accordingly.  Rays are deleted when they contain $< 10^{-10}\times$ the average number of photons per ray.  We use the full speed of light to maintain accuracy at the end of reionization.  

As the rays propagate, they adaptively split to maintain a minimum angular resolution around the source cell.  When rays from many sources intersect the same cell, the ones with the fewest photons are merged to a fixed level of angular resolution.  Splitting and merging is handled with the HealPix formalism~\citep{Gorski1999} following a procedure similar to the one described in~\citet{Abel2002} and implemented in~\citet{Trac2007}.\footnote{In fact, we have tested our code against that of~\cite{Trac2007} and found excellent agreement in the shapes and sizes of ionized and neutral regions.}  The parameters for this are adjustable, allowing the user to trade accuracy for computational time.  In Appendix~\ref{app:conv}, we describe these parameters and show that our choices for them are converged in terms of morphology.  

To maximize flexibility, our RT algorithm does not explicitly solve for the ionization state of each cell to determine its absorption coefficient, $\overline{\kappa}$.  Instead, $\overline{\kappa}$ can be an arbitrary function of density, photo-ionization rate, ionization redshift, and time.  Moreover, since our RT cells are large enough to require many RT steps to ionize (1 $h^{-1}$Mpc in this work), we track the I-fronts within cells using a ``moving screen'' approximation.  That is, I-fronts are infinitely sharp and the gas behind them is highly ionized.  The photo-ionization rate in ionized gas is given by
\begin{equation}
    \label{eq:gammah1_subgrid}
    \Gamma_{\rm HI}^{i} = \sum_{j=1}^{N_{\rm rays}}\frac{N_{\gamma,0}^{ij} \overline{\sigma}_{\rm HI} \overline{\lambda}^{i} [1 - \exp(-x_{\rm ion}^{i} \Delta s^{ij}/\overline{\lambda}^{i})]}{x_{\rm ion}^{i} V_{\rm cell}^{i} \Delta t},
\end{equation}
where the number of photons in ray $j$ traveling a distance $\Delta s_{ij}$ through cell $i$ is $N_{\gamma,0}^{ij}$, $\overline{\lambda}^{i} \equiv 1/\overline{\kappa}^{i}$ is the mean free path, $x_{\rm ion}^{i}$ is the ionized fraction, $V_{\rm cell}$ is the cell volume, and the sum is over all rays crossing cell $i$ during the time step $\Delta t$.  The cross-section $\overline{\sigma}_{\rm HI}$ is averaged over the assumed spectrum of $J_{\nu} \propto \nu^{-1.5}$ from $1-4$ Ryd (as in~\citealt{DAloisio2020}, motivated by the scaling anticipated in stellar population synthesis models).  In partially ionized cells, I-fronts move at a speed $v_{\rm IF} = {F_{\gamma}}/[{(1+\chi)n_{\rm H}}]$, where $\chi = 0.082$ accounts for HeI and $F_{\gamma}$ is the leftover photon flux after attenuation by the ionized part of the cell.  
In Appendix~\ref{app:gamma_deriv} we show explicitly that Eq.~\ref{eq:gammah1_subgrid} is valid for arbitrary $\overline{\kappa}$.  

\subsection{Sub-grid model for $\overline{\lambda}$} 
\label{subsec:subgrid}

In standard RT, Eq.~\ref{eq:gammah1_subgrid} would be closed by an ionization balance equation (perhaps including a sub-grid clumping factor) and $\overline{\lambda}$ computed from the HI fraction.  Our fiducial setup instead uses a prescription for $\overline{\lambda}$ based on an extended suite of the small-volume hydro plus ray-tracing RT simulations first presented in~\citet{DAloisio2020}.  These were run with a modified version of the RadHydro code~\citep{Trac2004, Trac2007} in $1$ (Mpc/h)$^3$ volumes with $N = 1024^3$ DM particles, gas and RT cells.  
We ionize the whole volume at $z = z_{\rm reion}$ by sending I-fronts from the boundaries of $L_{\rm dom} = 32$ $h^{-1}$kpc domains.  This setup avoids complicating the interpretation of our results with uncertain galaxy physics by treating the gas as if it were reionized by external sources.  The photo-ionization rate $\Gamma_{\rm -12} \equiv \Gamma_{\rm HI}/(10^{-12} \text{ s}^{-1})$ is constant in optically thin gas. (We emphasize, however, that our simulations explicitly include self-shielding systems and associated RT effects.)  We simulated over-dense and under-dense regions by using the method of~\citet{Gnedin2011} to account for box-scale density fluctuations.  These are parameterized by $\delta/\sigma$, the linearly extrapolated over-density in units of its standard deviation.  We refer the reader to~\citet{DAloisio2020} for more details\footnote{Our expansion of the suite in~\citet{DAloisio2020} includes all combinations of $z_{\rm reion} \in \{6, 8, 12\}$, $\Gamma_{-12} \in \{0.03, 0.3. 3.0\}$ and $\delta/\sigma \in \{-\sqrt{3}, 0, \sqrt{3}\}$.  Due to computational limitations, not all of our small-volume simulations are run to when reionization ends ($5 < z < 6$).  In these cases we extrapolate the results to lower redshifts by fitting $\overline{\lambda}$ to a power law in cosmic time over the last $50$ Myr of the run.  }.  

We estimate $\overline{\lambda}$ in our RadHydro simulations using
\begin{equation} 
    \label{eq:lambda_freq}
    \overline{\lambda}^{-1} \equiv \overline{\kappa} = \frac{\langle \Gamma_{\rm HI} n_{\rm HI}\rangle_{\rm V}}{F_{\gamma}},
\end{equation}
where $F_{\gamma}$ is the ionizing photon flux in each domain.  
In Appendix~\ref{app:lambda_deriv} we show that the right-hand side of Eq.~\ref{eq:lambda_freq} is equal to the volume-averaged absorption coefficient and is the relevant quantity for evaluating Eq.~\ref{eq:gammah1_subgrid}.  Note that this definition of $\overline{\lambda}$ accounts for non-equilibrium absorptions by self-shielded systems (e.g. mini-halos), an effect that cannot be accurately captured with a clumping factor~\citep[as noted by][]{McQuinn2007,Shukla2016}.  

Our RadHydro simulations give us $\overline{\lambda}$ versus time in a range of environments parameterized by $(z_{\rm reion}, \Gamma_{\rm HI}, \delta/\sigma)$.  While we could simply interpolate over these parameters to get $\overline{\lambda}^i$ in Eq.~\ref{eq:gammah1_subgrid}, doing so would neglect the sensitivity of $\overline{\lambda}$ to the time-evolution of $\Gamma_{\rm HI}$, since $\Gamma_{\rm HI}$ does not evolve in the small-volume simulations.  This sensitivity arises from the dependence of the relaxation process on the self-shielding properties of the gas, which are set by largely by $\Gamma_{\rm HI}$ (see Figs. 5 and 6 of~\citealt{DAloisio2020}).  We incorporated this $\Gamma_{\rm HI}$-dependence using an empirically-motivated model for the full time-evolution of $\overline{\lambda}$, 
\begin{equation}
	    \label{eq:lambdamaster}
	    \frac{d \overline{\lambda}}{dt} = \pd{\overline{\lambda}}{t}\Big|_{\Gamma_{\rm HI}} + \pd{\overline{\lambda}}{\Gamma_{\rm HI}} \Big|_{t}\frac{d\Gamma_{\rm HI}}{dt} - \frac{\overline{\lambda} - \overline{\lambda}_0}{t_{\rm relax}},
\end{equation}
where the first term captures the time-dependence of $\overline{\lambda}$ at fixed $\Gamma_{\rm HI}$, and the second the instantaneous change in $\overline{\lambda}$ with $\Gamma_{\rm HI}$.  The former is interpolated from our small-volume simulation suite, and for the latter we assume a power law $\overline{\lambda} \propto \Gamma_{\rm HI}^{2/3}$, consistent with the scaling found in simulations~\citep[e.g.][]{McQuinn2011}. The last term captures the evolution of $\overline{\lambda}$ towards the constant-$\Gamma_{\rm HI}$ limit $\overline{\lambda}_0$ (also interpolated from our small-volume suite).  Here $t_{\rm relax}$ is the timescale over which the gas loses memory of its previous $\Gamma_{\rm HI}$ history, which we take to be $100$ Myr.  In Appendix~\ref{app:evolving_gamma}, we show that Eq.~\ref{eq:lambdamaster} compares well against small-volume simulations with evolving $\Gamma_{\rm HI}$.  Since $\overline{\lambda}$ is a function of $\Gamma_{\rm HI}$, Eqs.~\ref{eq:gammah1_subgrid}  and~\ref{eq:lambdamaster} are iterated five times for each time step, which we find sufficient for convergence (Appendix~\ref{app:conv}).  

\subsection{Caveats}
\label{subsec:caveats}

Here we will briefly discuss two caveats to our sub-grid model.  The first is that our small-volume simulations should under-produce massive halos, which can act as sinks.  This may be true even in our over-dense DC mode runs, which sample biased regions of the IGM where these halos are more common.  This would be most problematic at the lowest redshifts when rare, massive sinks contribute significantly to the IGM opacity~\citep{Nasir2021}\footnote{In~\citet{Cain2021}, this issue partially motivated the enhanced sinks model, which appealed to missing rare sinks to help explain the mild evolution of the mean free path at $z < 5$.  }.  

The second concerns our treatment of self-shielded gas.  Eq.~\ref{eq:lambda_freq} for $\overline{\lambda}$ accounts for absorptions by self-shielded gas clumps that remain neutral some time after I-front passage~\citep{Nasir2021}.  The gas in these systems can be a significant fraction of the gas in the cell within $50$ Myr of ionization when $\Gamma_{\rm HI}$ is low ($\lesssim 10^{-13}$ s$^{-1}$).  In principle, this gas should be excised from our moving-screen I-front calculation, which counts $1$ absorption per neutral atom during I-front passage.  As such, gas that remains neutral for more than a few Myr after I-front passage is effectively treated as if it were ionized twice.  We have run a conservative test in which we derive $\overline{\lambda}$ in the small-volume simulations using the recombination clumping factor $C_R$ (Eq. 5 of~\citealt{DAloisio2020}) under the assumption of photo-ionizational equilibrium.  This approach ignores the fact that some of the neutral gas is ionized after I-front passage and counts only recombination-balanced absorptions (see \S\ref{subsec:hydrovis} in the next section for more details).    Thus using $C_R$ likely under-estimates the photon budget and brackets the magnitude of the double-counting effect.
We found that the difference between the number of absorptions in ionized gas between using $C_R$ and our fiducial model can be as high as a factor of $2$ when low-$\Gamma_{\rm HI}$ gas dominates the absorption rate.  Thus the photon budget predicted by our fiducial sinks model is almost certainly too high, although which model is closer to the truth is unclear.  Fortunately, the impact on our results is minimal because, as we will see, the sinks probably do not shape morphology substantially under most circumstances.  Even so, our results using this model should be interpreted as an upper limit on the expected effect of un-relaxed gas.  In what follows we will make note whenever this point becomes relevant.  

\subsection{Density Fields \& Sources} \label{subsec:sources}

The density and source fields for our large-volume RT simulations are taken from a cosmological N-body DM-only simulation in a $300$ $h^{-1}$Mpc box run using MP-Gadget~\citep{Feng2018}.  The run used $N = 2048^3$ DM particles, for a mass resolution of $2.5 \times 10^8$ $h^{-1}$M$_{\odot}$ and a minimum halo mass of $8.5 \times 10^9$ $h^{-1}$M$_{\odot}$ (corresponding to $32$ DM particles).  The DM particles were smoothed onto a grid with $1$ $h^{-1}$Mpc cells to get the density fields for the RT calculation.  Density and halo fields are updated every $10$ Myr from $z = 12$ to $4.5$, for a total of $99$ snapshots.  The RT time-step is equal to the light-crossing time of the RT cells, and varies from $\approx 0.4$ to $\approx 0.8$ Myr during the simulation.  When the density field is updated, we keep the same ionized fractions in all cells - thus we neglect the advection of ionized/neutral gas between snapshots.  This should be a reasonable approximation since bulk velocities on $\geq 1$ $h^{-1}$Mpc scales are typically slower than the speed of ionization fronts (a few hundred vs. $10^3 - 10^4$ km/s).  We assigned UV luminosities to halos by abundance matching to the UV luminosity function of~\citet{Finkelstein2019}.  

Halos with masses well below $8.5 \times 10^9$ $h^{-1}$M$_{\odot}$ likely formed stars via atomic cooling, and so may have contributed significantly to reionization.  We thus extended the halo mass function (HMF) of our simulation using a modified version of the non-linear biasing method of~\citet{Ahn2015}.  These ``sub-resolved'' halos follow the HMF of~\citet{Watson2013} (which agrees with our resolved HMF) and are spatially distributed following the extended Press-Schechter (EPS) formalism.  The number of added halos in each cell and mass bin is drawn randomly from a Poisson distribution with mean equal to the halo abundance predicted by EPS.  We found that the clustering of the halos predicted by this formalism was systematically higher than that in the SCORCH simulations~\citep{Trac2015}.  Specifically, the halo bias produced by the EPS method was a factor of $1.4$ $(1.32, 1.18)$ too high compared to SCORCH at $z = 10$ $(8, 6)$.  We therefore added an empirically derived bias correction to the model to approximately reproduce the clustering of SCORCH halos in the mass range of interest.

We extended the HMF in our simulations to a minimum mass of $M_{\min} = 10^9$ $h^{-1} M_{\odot}$.  Emissivities were assigned to halos assuming that the emissivity of each halo follows a power law in UV luminosity, $\dot{n}_{\gamma} \propto L_{\rm UV}^{\beta}$.  Smaller $M_{\min}$ and $\beta$ correspond to reionization driven by fainter, less biased sources.  Our fiducial model has $M_{\min} = 10^9$ $h^{-1} M_{\odot}$ and $\beta = 1$, which corresponds to assuming a single value of the escape fraction $f_{\rm esc}$ and ionizing efficiency $\xi_{\rm ion}$ for the entire source population at each redshift.  We chose this as our fiducial model for two reasons: (1) it imposes minimal assumptions about the dependence of $f_{\rm esc}$ and $\xi_{\rm ion}$ on halo mass and (2) of the models we will consider, it is the most similar to models commonly used in reionization simulations (e.g. $n_{\gamma} \propto M$ as in~\citet{Keating2019} and \citet{Mao2019}).  In \S\ref{sec:sources} we study what happens when $M_{\min}$ and $\beta$ are varied.  In all simulations, the global emissivity (summed over all halos) as a function of redshift is an input chosen to produce the desired reionization history.  Our models all use re-scaled versions of the fiducial late-ending rapid model of~\citet{Cain2021}, as shown in the middle panel of Figure~\ref{fig:ion_history_sinks}.

One caveat of this method is that the sub-resolution halos (with $M < 8.5 \times 10^9~h^{-1}$ M$_\odot$) that are added, being randomly drawn at each 10 Myr time-step, are not causally connected -- i.e. halos jump around between time steps.  This is an insignificant effect in over-dense regions containing many halos, where the ``shot noise" is small, but can be pronounced in under-dense regions containing very few halos.  We have run a series of tests against idealized scenarios in which the halos are held in fixed locations throughout reionization.  We find that the noise introduced by the random drawing tends to wash out the smallest structures in the ionization field, but that on the larger scales of interest the effects are modest.  In general, we found slightly {\it less} power in the ionization field on large scales ($k \lessapprox 0.5$ $h$Mpc$^{-1}$) in our ``fixed sources'' tests.  We find that the effect is never large enough to affect any of our forthcoming results at the qualitative level.  We will discuss quantitative details of these tests in the results sections when they become relevant.

\section{The Effect of Sinks on Reionization's Morphology}
\label{sec:results}

\subsection{Sinks Models}
\label{subsec:sinksmodels}

\begin{figure*}
    \centering
    \includegraphics[scale=0.155]{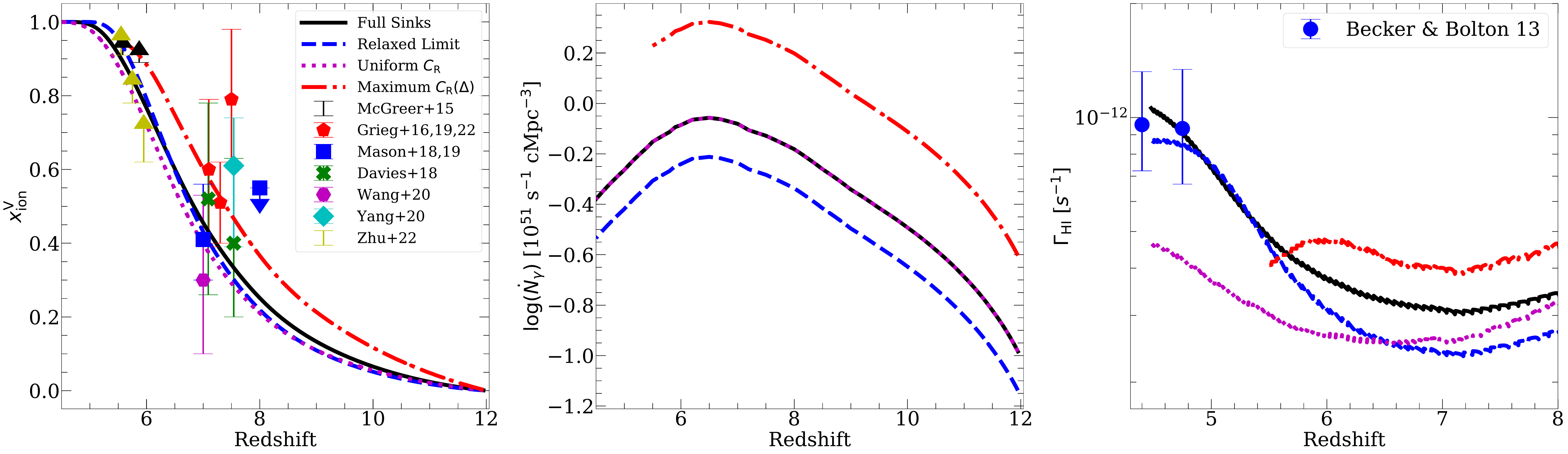}
    \caption{Volume-averaged ionized fraction (left), co-moving total ionizing emissivity (middle), and average photo-ionization rate in fully ionized cells (right) for each sinks model.  All results shown here adopt our fiducial source scenario.  We include measurements from the literature in the left panel \citep{McGreer2015, Grieg2016,Grieg2019,Davies2018,Mason2018, Mason2019, Wang2020,Yang2020a,Grieg2022,Zhu2022}.  All the reionization histories are similar except the \maxC~case. In the ensuing discussion, we show that the {\UrlFont Full Sinks} and \unifC~models have nearly indistinguishable morphologies assuming our fiducial source model. Notably, although these models have the same reionization and emissivity histories, they have significantly different photo-ionization rates. Hence they may be distinguishable by observables that are sensitive to $\Gamma_{\rm HI}$, e.g. the mean free path and the Ly$\alpha$ forest.  }
    \label{fig:ion_history_sinks}
\end{figure*}

In this section, we discuss the effect of sinks on the morphology of reionization.  We compare our new sinks model to several representative alternatives.  We assume our fiducial source model throughout (in \S\ref{sec:sources} we will explore others.)  We compare the following sink prescriptions: 

\begin{itemize}
    \item {\bf \UrlFont Full Sinks}: Our fiducial sinks model is based on the suite of RadHydro simulations as described in \S~\ref{subsec:subgrid}.  The evolution of $\overline{\lambda}$ in each cell includes the dynamical effects of pressure smoothing and photoevaporation, as well as the impact of sub-resolved self-shielding on the IGM opacity.  
    
    \item {\bf \UrlFont Relaxed Limit}: For this model, we extrapolate the low-redshift $\overline{\lambda}$ from our $z_{\rm reion} = 12$ RadHydro simulations to higher redshifts, assuming a power law in cosmic time, and directly interpolate $\overline{\lambda}$ instead of using Eq.~\ref{eq:lambdamaster}.  Thus, the gas is treated in the limit that it was ionized long ago and has reached a pressure-smoothed equilibrium.  This model effectively removes the contribution of opacity from the initial clumpiness that is eventually erased during the relaxation process.  
    
    \item {\bf Sub-grid Clumping Factor}: Here we assume that all gas in ionized regions is in photo-ionization equilibrium at a constant $T = T_{\rm ref} \equiv 10^4$ K, which yields
    \begin{equation}
        \label{eq:clumping_factor_model}
        \overline{\lambda} = \frac{\Gamma_{\rm HI}}{\overline{\sigma}_{\rm HI} C_{\rm R} \alpha_{\rm B}(T_{\rm ref}) (1+\chi) n_{\rm H}^2}
    \end{equation}
    where $\alpha_{\rm B}$ is the case B recombination coefficient of ionized hydrogen.  We adopt two prescriptions for $C_{\rm R}$: 
        \begin{enumerate}
            \item {\bf \unifCb}: We set $C_{\rm R} = 5$ everywhere at all times, which reproduces a reionization history and photon budget similar to the {\UrlFont Full Sinks} model.  This case serves as a basis for comparison to assess the importance of the dynamics and spatial in-homogeneity of the sinks predicted by the {\UrlFont Full Sinks} model.  We emphasize that $C_{\rm R}$ is a sub-grid clumping factor, not a global one.  
            
            \item {\bf \maxCb}: We use the density-dependent sub-grid clumping factor of~\citet{Mao2019}.\footnote{Note that the large-volume simulations in~\citet{Mao2019} have smaller cells than ours, so their clumping factors are a slight under-estimate for our application.  Still, this model serves the purpose of illustrating how the morphology evolves in an extremely clumpy IGM, which is our goal.  }  This model is based on dark-matter-only N-body simulations and predicts $C_{\rm R} \approx 10 - 15$ in cells with $\Delta \geq 1$ at $z \leq 8$. Since this model neglects pressure smoothing effects, it represents an upper limit on the amount of clumping in the standard cosmology.  
        \end{enumerate}
\end{itemize}

The left-most panel in Figure~\ref{fig:ion_history_sinks} shows the volume-averaged ionized fraction for each sinks model alongside measurements from the literature. The middle panel shows the global ionizing emissivity.  The emissivity histories are all re-scaled versions of the ``rapid" model from \citet{Cain2021}.  For comparison, the emissivities of the {\UrlFont Full Sinks}, {\UrlFont Relaxed Limit}, and \unifC~models have been tuned to yield very similar reionization histories and ionizing photon budgets, ending reionization late at $z = 5 - 5.5$.  The \maxC~emissivity was tuned to end reionization somewhat earlier because the clumping factor fits from~\citet{Mao2019} do not extend below $z \sim 6.5$\footnote{We extrapolate the~\citet{Mao2019} fitting parameters to slightly lower redshifts by assuming their $a_0$ parameter evolves linearly in redshift, while $a_1$ and $a_2$ retain their $z = 6.5$ values (see their Eq. 17 and appendix B.)  }.  However, the ensuing morphology comparisons will be performed at fixed global ionized fraction, which should minimize  any differences originating from the different reionization histories.  Note that the {\UrlFont Full Sinks} and \unifC~models have the same emissivity, while the {\UrlFont Relaxed Limit} (\maxC) emissivity is a factor of $0.7$ ($2.4$) smaller (larger) than the other two.  We note that due to the over-counting issue discussed in \S\ref{subsec:caveats}, the emissivity in the {\UrlFont Full Sinks} and \unifC~models are likely higher than they should be.  A lower photon budget would mean a smaller $C_{\rm R}$ in the latter to match the {\UrlFont Full Sinks} case; thus the value of $C_{\rm R} = 5$ is probably too high.  In the ensuing discussion we will see that our main conclusions on morphology are not significantly affected by this issue.  

The right-most panel of Figure~\ref{fig:ion_history_sinks} shows $\Gamma_{\rm HI}$ averaged over fully ionized cells for each model, compared to measurements from~\citet{Becker2013}. Here we omit $z>5$ measurements \citep[e.g.][]{Calverley2011,Wyithe2011,DAloisio2018,Becker2021} for clarity, and also because it is unclear how to compare these measurements against our $\Gamma_{\rm HI}$ in simulations where reionization is still ongoing at $z=5-6$.    A number of reionization observables are explicitly sensitive to $\Gamma_{\rm HI}$, including the mean free path, Ly$\alpha$ forest statistics, and LAE visibility. In the ensuing discussion we will show that the {\UrlFont Full Sinks} and \unifC~models exhibit essentially identical morphologies in our fiducial source model. A key takeaway from Figure \ref{fig:ion_history_sinks} is that sink models tuned to yield similar morphologies, e.g. the {\UrlFont Full Sinks} and \unifC~models, may nonetheless exhibit considerable differences in observables that are sensitive to $\Gamma_{\rm HI}$.  So while these models may appear nearly identical in their predictions for the 21cm power spectrum, they will yield different predictions for e.g. Ly$\alpha$ forest statistics.  

\subsection{Visualization of the IGM Opacity}
\label{subsec:hydrovis}

To aid in visualizing the dynamics and spatial morphology of the sinks, we define the ``effective clumping factor'' for cell $i$ to be
\begin{equation} 
\label{eq:clumpeff}
    C_{\rm eff}^{i} = \frac{(1/\overline{\lambda}^{i})}{\overline{\sigma}_{\rm HI}  \alpha_B(T_{\rm ref}) (1+\chi) {n_{\rm H}^{i}}^2/\Gamma_{\rm HI}^{i}},
\end{equation}
where $T_{\rm ref} = 10^4$ K and $\Gamma_{\rm HI}^{i}$, $\overline{\lambda}^{i}$, and $n_{\rm H}^{i}$ are the photo-ionization rate, mean free path, and H number density, respectively.  The numerator is simply the absorption coefficient $\kappa$, and the denominator is what $\kappa$ would be if the gas had a constant temperature $T_{\rm ref}$, was in photo-ionizational equilibrium, and had no sub-resolved density fluctuations.  $C_{\rm eff}$ quantifies the impact of sub-grid sink physics and large-scale temperature fluctuations on the opacity.  In the limit of photo-ionizational equilibrium, Eq.~\ref{eq:clumpeff} is equivalent to the recombination clumping factor $C_R$ (see \S~\ref{subsec:caveats}).  Differences between $C_{\rm eff}$ and $C_R$ indicate the presence of sub-resolved self-shielded systems that are not in photo-ionizational equilibrium.  Note that unlike in~\citet{DAloisio2020}, the density in the denominator of our clumping factors is the {\it cell-wise} density rather than the cosmic mean density.  Thus, density fluctuations influence $C_{\rm eff}$ only indirectly through their impact on the sub-resolved clumpiness of the gas and its self-shielding properties.   

\begin{figure}
    \centering
    \includegraphics[scale=0.17]{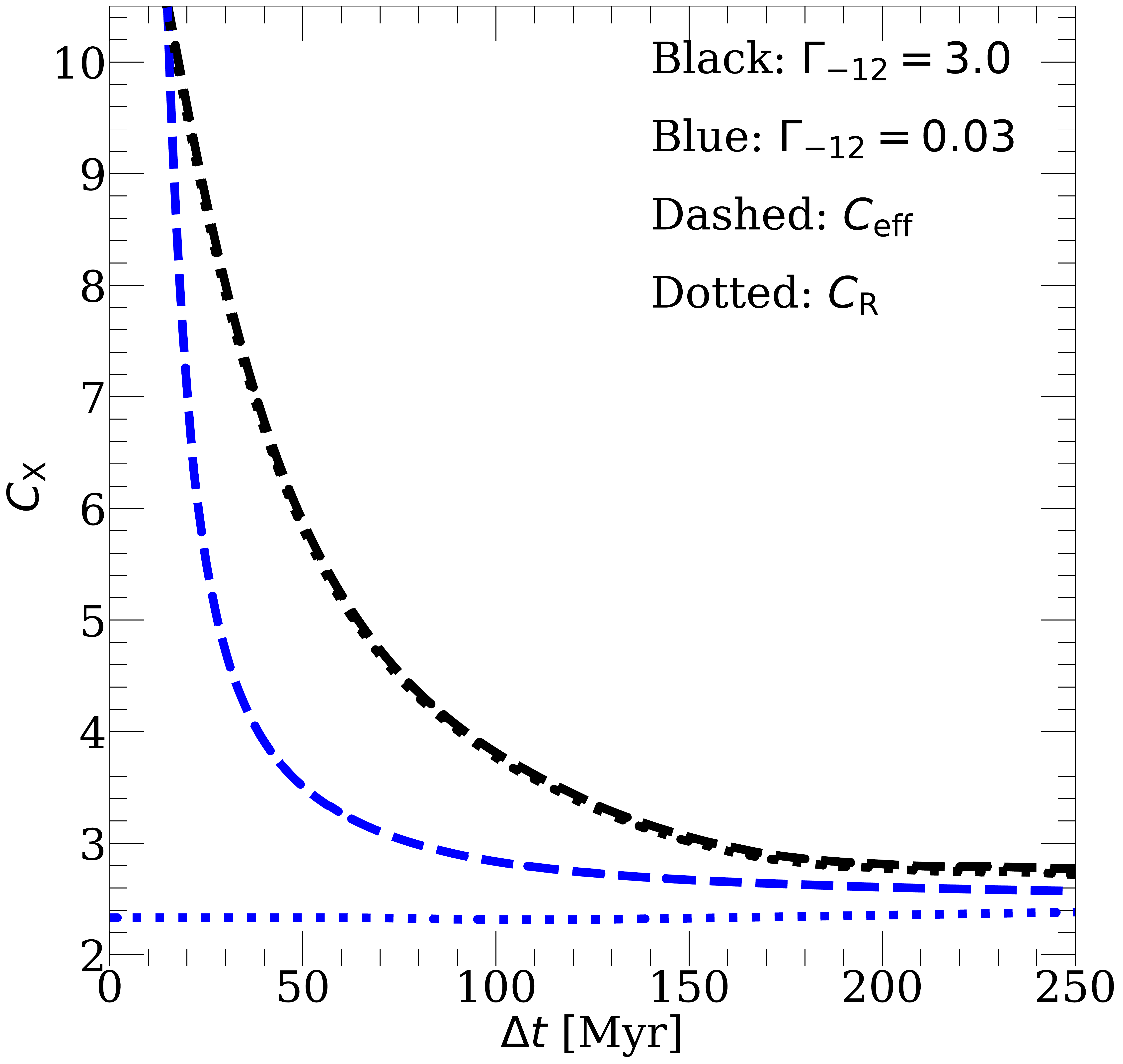}
    \caption{Examples of $C_{\rm eff}$ (dashed) compared to $C_R$ (dotted) for small-volume simulations with high and low values of $\Gamma_{-12}$ ($3.0$ and $0.03$).  In the former, the two quantities are similar owing to the scarcity of self-shielded gas.  However for $\Gamma_{-12} = 0.03$, systems remain self-shielded and out of equilibrium for longer, producing a large difference between $C_{\rm eff}$ and $C_R$ since the former reflects the total number of absorptions but the latter only those balanced by recombinations.  }
    \label{fig:radhydro_subgrid_clumpeff_example}
\end{figure}

\begin{figure*}
    \centering
    \includegraphics[scale=0.074]{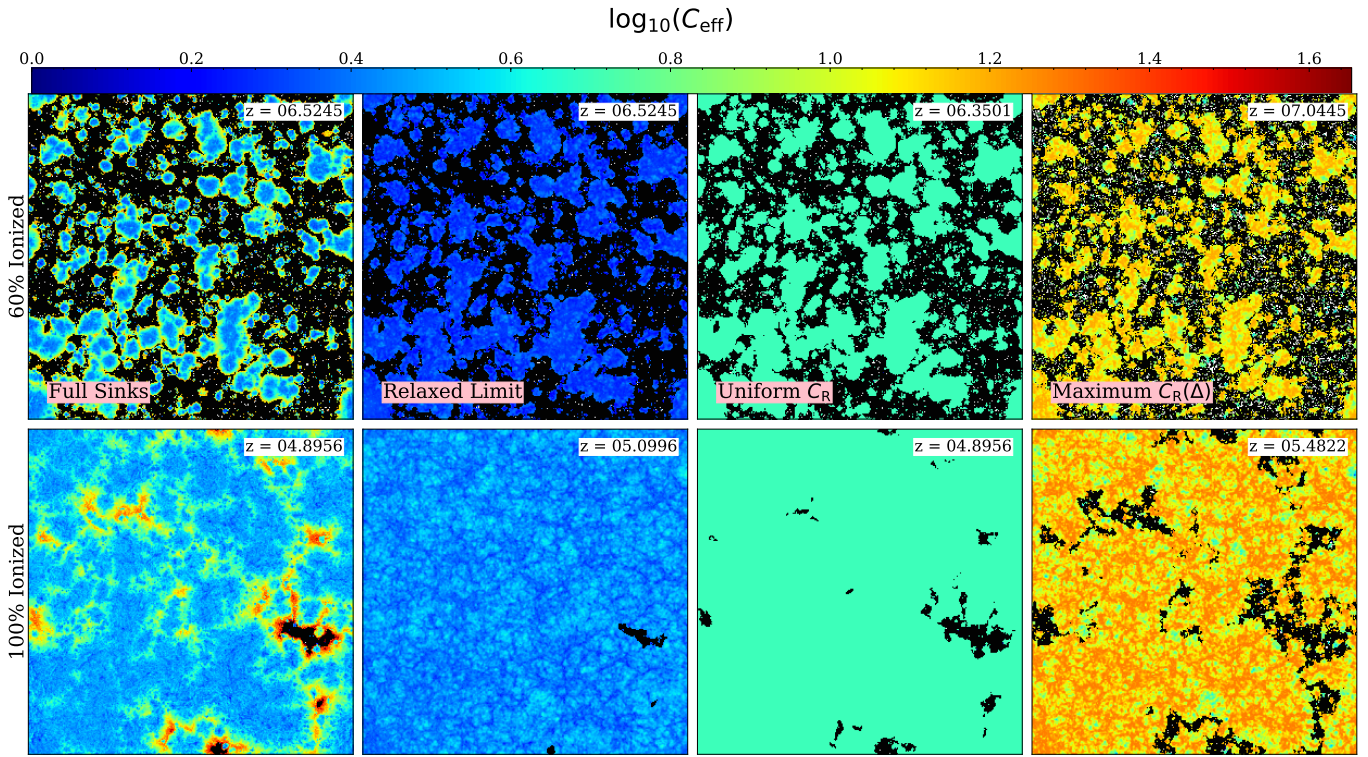}
    \caption{Visualization of the sink physics in each of our models.  The redshifts are notated in the upper right of each panel.  We show $\log_{10}(C_{\rm eff})$ at 60\% volume ionized fraction (top) and $50$ Myr after reionization ends ($x_{\rm ion}^{\rm V} < 0.01$, bottom row).  The black regions denote cells that are at least $50\%$ ($10\%$) neutral in the top (bottom) panels.  In the {\UrlFont Full Sinks} case, the opacity is boosted near I-fronts (top) and in under-dense voids that have yet to relax after reionization ends (bottom).  The large scale fluctuations in $C_{\rm eff}$ are weaker in the other models.  In the \unifC~case, $C_{\rm eff}$ is the same everywhere, and in the \maxC~models, $C_{\rm eff}$ is lower (higher) than average in voids (over-densities) after reionization, in contrast to the {\UrlFont Full Sinks} case. These visualizations illustrate the dynamical effects of pressure smoothing and photoevaporation in our {\UrlFont Full Sinks} model.  }
    \label{fig:relaxation_vis}
\end{figure*}

For intuition on $C_{\rm eff}$, Figure~\ref{fig:radhydro_subgrid_clumpeff_example} shows its evolution (dashed curves) compared to that of $C_R$ (dotted curves) vs. time since ionization for two of our mean density, $z_{\rm reion} = 8$ small-volume RadHydro simulations.  One has $\Gamma_{-12} = 3.0$ (black), and the other $0.03$ (blue).  In the first case, $C_{\rm eff}$ and $C_R$ are close together; both start above $10$ and approach $\sim 3$ as the gas relaxes.  Their similarity owes to the high intensity of the background, which leaves little gas self-shielded.  In the $\Gamma_{-12} = 0.03$ case, there is significant self-shielding in high-density gas.  This lowers $C_R$ (which counts only recombination-balanced absorptions), while $C_{\rm eff}$ remains elevated, since it is affected by non-equilibrium absorptions taking place as the self-shielded gas is ionized.  At later times, $C_{\rm eff}$ and $C_R$ agree better as more self-shielded systems evaporate.

Figure~\ref{fig:relaxation_vis} shows slices of $\log(C_{\rm eff})$ from large-volume simulations for each of our sinks models (assuming our fiducial source model).  We show the {\UrlFont Full Sinks} (left-most), {\UrlFont Relaxed Limit} (middle left), \unifC~(middle right) and \maxC~(right-most) models at $60$\% volume ionized in the top row, and $50$ Myr after reionization has finished ($x_{\rm ion}^{\rm V} = 0.01$) in the bottom row.  The redshifts are given in the upper right of each panel.  In the top (bottom) row, black regions denote cells that are at least $50\%$ ($10\%$) neutral (note that a small number of cells are still partially neutral even after $x_{\rm ion}^{\rm V} < 0.01$ in the bottom row).  In the {\UrlFont Full Sinks} model, $C_{\rm eff}$ is highest near I-fronts where gas was most recently ionized.  After reionization ends, patches of enhanced opacity with $C_{\rm eff} \sim 10-20$ (and even higher in the most recently ionized cells) persist in the voids, which ionized last and quickly, so have yet to relax.  In regions re-ionized earlier, $C_{\rm eff}$ is $\sim 2 - 5$ at all redshifts, similar to the {\UrlFont Relaxed Limit}.  The opacity is higher in the \unifC~case than in the {\UrlFont Relaxed Limit} because it has been calibrated to match the photon budget of the {\UrlFont Full Sinks} model.  The \maxC~model has the highest opacity, with $C_{\rm eff} \sim 10-20$ everywhere after reionization.  

A comparison between the top-left and the two top-right panels in Fig. \ref{fig:relaxation_vis} reveals that the opacity in over-dense regions hosting the earliest ionized bubbles is significantly lower in our {\UrlFont Full Sinks} model compared to the \unifC~and \maxC~models.  This results from the dynamics in our {\UrlFont Full Sinks} model, and may arise from two effects working in tandem: (1) $\Gamma_{\rm HI}$ is generally larger near the highly clustered sources, which leads to a quicker relaxation/evaporation of the sinks nearby ; (2) The structures that form in these regions may have a shorter relaxation time owing to their larger densities \citep[see e.g. Eq. 4 of][]{DAloisio2020}.  Together, these effects in our {\UrlFont Full Sinks} model work towards favoring the growth of larger bubbles compared to the \unifC~and \maxC~models. Conversely, the opacity is elevated in recently ionized regions at lower redshifts near the end of reionization, despite these regions being under-dense on average.

In the other three models, $C_{\rm eff}$ is affected mainly by density fluctuations, which are most noticeable in the \maxC~model (and absent by construction in the \unifC~case). 
In the \unifC~model all parts of the IGM have the same $C_{\rm R}$, while in the \maxC~model the over(under) dense regions have the highest (lowest) $C_{\rm R}$, opposite the {\UrlFont Full Sinks} case.
We emphasize that the contrasting $C_{\rm eff}$ topologies will affect any observables that are explicitly sensitive to $\Gamma_{\rm HI}$ and the opacity structure of the IGM, such as the Ly$\alpha$ forest and the mean free path (see discussion of Fig.~\ref{fig:ion_history_sinks}).  However, in the ensuing discussion we will see that they are probably not very important for morphology.  

\subsection{Ionized Bubbles}
\label{subsec:ionbubbles}

\subsubsection{Visualization of Ionized Region Morphology}
\label{subsubsec:ionvis}

\begin{figure*}
    \centering
    \includegraphics[scale=0.141]{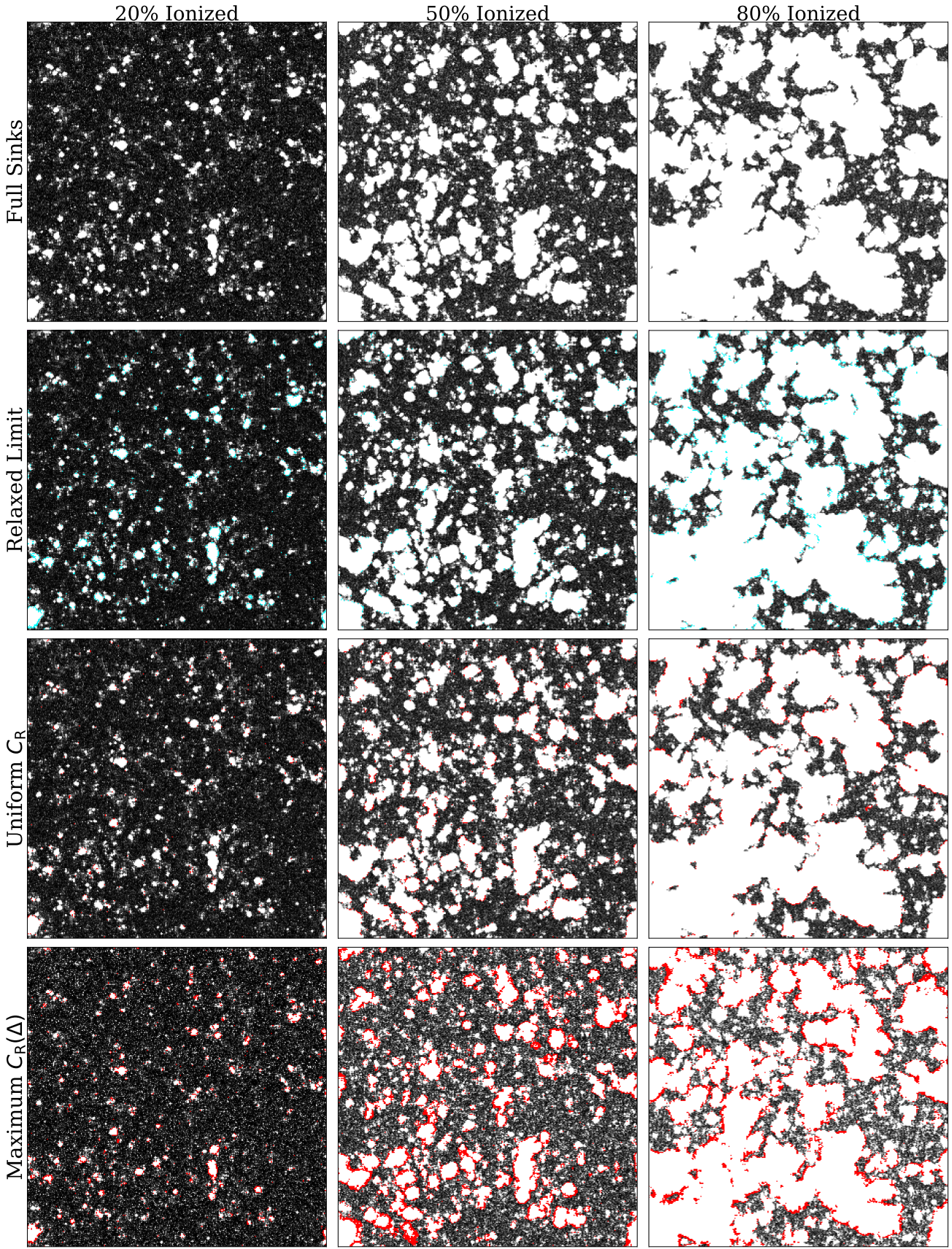}
    \vspace{-0.1cm}
    \caption{Visualization of the ionization field for our sinks models. All results here correspond to our fiducial source model with $\dot{n}_\gamma \propto L_{\rm UV}$, i.e. assuming the same escape fraction and ionizing efficiency for all sources.  The columns show different volume ionized fractions ($20$, $50$, and $80$\%, left to right) and the rows show different sinks models.  In the second row, the cyan shading indicates bubbles that are slightly larger than in the {\UrlFont Full Sinks} model, while the red shading indicates the opposite in the other two rows.  The ionized bubbles are smallest in the \maxC~model at all ionized fractions.  The {\UrlFont Relaxed Limit} model has slightly larger bubbles than the {\UrlFont Full Sinks} and \unifC~models, but these three models are visually very similar.  }
    \label{fig:ion_vis_sinks}
\end{figure*}

Figure~\ref{fig:ion_vis_sinks} shows the ionization field (darker = more neutral) for each of our sinks models (top to bottom, see labels) at 20, 50, and 80\% volume ionized fraction (left to right).  At fixed ionized fraction, the \maxC~model exhibits the smallest ionized bubbles.  This is indicated by the red shading, which denote regions that are neutral in the \maxC~(and \unifC) model, but not the {\UrlFont Full Sinks} case.  The other models are visually similar to the {\UrlFont Full Sinks} case - the \unifC model having slight smaller bubbles and the {\UrlFont Relaxed Limit} model having slightly larger ones (as indicated by the cyan shading in that row).  The largest bubbles are smaller in the \maxC~model because the sources driving their growth are ``taxed'' disproportionately by recombinations compared to those in smaller bubbles~\citep[][]{Furlanetto2005}.\footnote{This has been termed ``taxing  the rich" by \citet{Furlanetto2005}.}  Since large bubbles form in over-densities and start growing the earliest, their growth is slowed by recombinations sooner than their later-forming counterparts inhabiting lower densities.  Thus the sinks act to reduce the average bubble size at fixed ionized fraction~\citep[as found by e.g.][]{Furlanetto2005,McQuinn2007,Alvarez2012,Mao2019,Chen2022}.

Comparing the {\UrlFont Full Sinks} (top row) and \maxC~(bottom row) models, the ionized bubbles generally appear larger in the former at fixed ionized fraction.  As described in the previous section, this is a direct result of the dynamics in our sub-grid sinks model.  In the earliest bubbles to form around highly clustered sources, the sinks relax/evaporate quickly, allowing the bubbles to grow more easily. By contrast, the smaller bubbles that start growing around less biased sources generally encounter a clumpier IGM for longer periods of time. Together, these effects work toward favoring the growth of large bubbles and partially cancel the ``taxing the rich" effect described in the previous paragraph.  The \maxC~model instead has higher clumping factors at higher densities, which slows the growth of the largest bubbles more. In other words, our {\UrlFont Full Sinks} model taxes the rich {\it less} than the \maxC~model, which does not include any dynamical effects.      

Interestingly, in Figure~\ref{fig:ion_vis_sinks} we see a striking degree of similarity between the {\UrlFont Full Sinks} and \unifC~models at all ionized fractions.  In fact, these models do not even differ significantly from the {\UrlFont Relaxed Limit} except near the beginning of reionization.    The visual similarity leads us to one of our key conclusions, which we will hash out quantitatively in the ensuing sections.   Accounting for the pressure-smoothing of the IGM by reionization is crucial for modeling morphology accurately.   However, as long as this effect is accounted for ``on average,'' e.g. in the simplest case with a uniform sub-grid clumping factor, the detailed dynamics and spatial in-homogeneity of the sinks are likely not very important for morphology. We emphasize, however, that this conclusion holds only for source models in which bright galaxies contribute significantly to the ionizing photon budget, as in our fiducial source model. In \S\ref{sec:sources}, we will see scenarios for which the details of the sink modeling {\it do} become quite important. 

\subsubsection{Bubble Size Distribution}
\label{subsubsec:bubble}

\begin{figure*}
    \centering
    \includegraphics[scale=0.16]{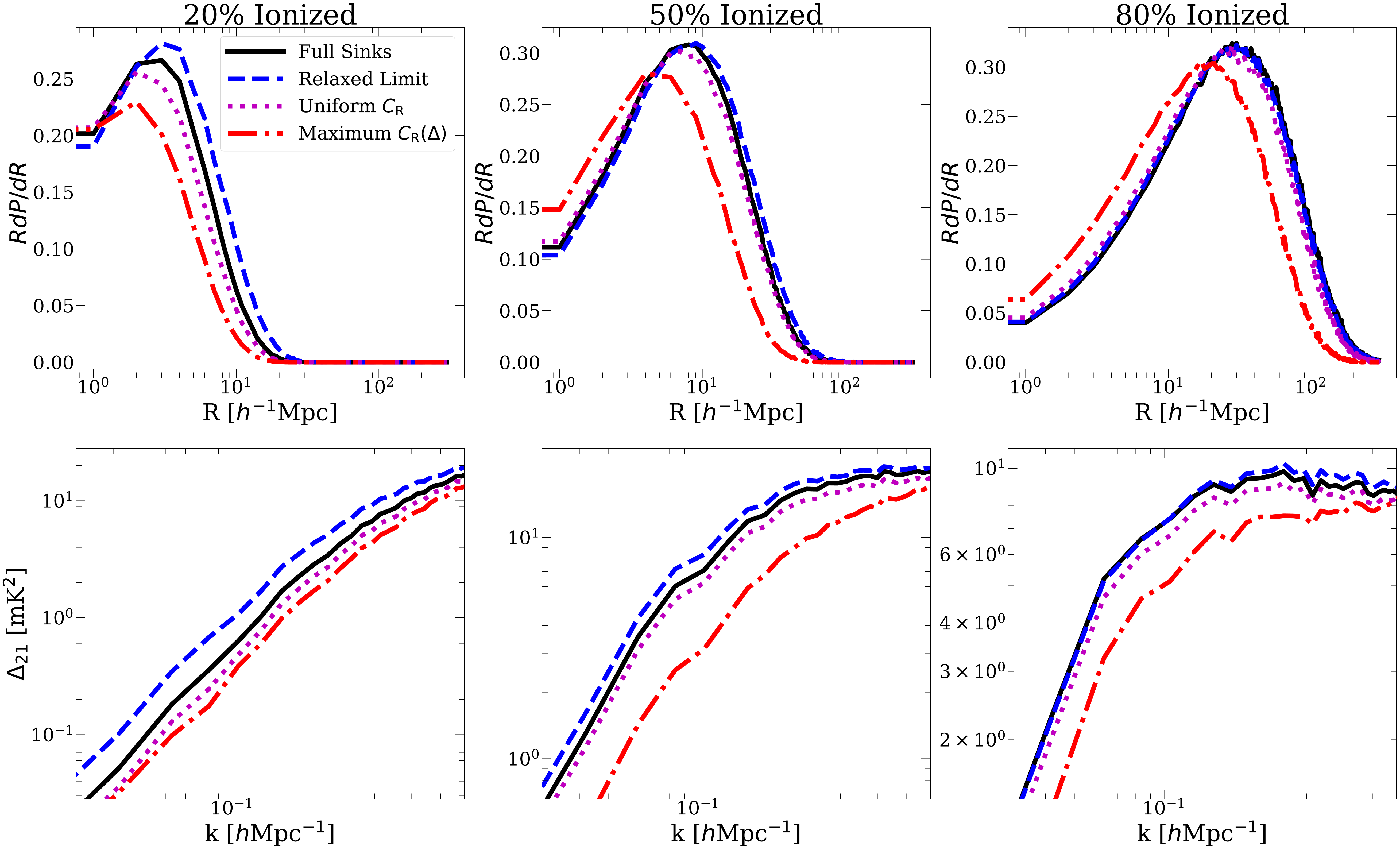}
    \caption{{\bf Top}: Ionized Bubble Size Distribution (IBSD) for our sinks models (see legend) at 20\%, 50\%, and 80\% volume ionized fractions (left to right). The {\UrlFont Full Sinks} and \unifC~models have strikingly similar IBSDs despite their very different clumping topologies.  {\bf Bottom}: 21 cm power ($\Delta_{\rm 21}$) vs. wavenumber for the same models and ionized fractions.  At $20\%$ ionized, the \maxC~model is well below the {\UrlFont Relaxed Limit},  with the other two models in between, but closer to the \maxC~result.  At later times, the {\UrlFont Full Sinks} and \unifC~models (which are always very similar to each other) move close to the {\UrlFont Relaxed Limit}. All results shown here adopt our fiducial source model with $\dot{n}_\gamma \propto L_{\rm UV}$.  }
    \label{fig:bubble_sinks}
\end{figure*}

Next, we study morphology more quantitatively using the ionized bubble size distribution (IBSD).  We compute the IBSD using the ray-tracing definition proposed in~\citet{Mesinger2007} and implemented in the publicly available package {\it tools21cm}~\citep{Giri2018}.  The IBSD defined this way captures the distribution of distances to neutral gas along random rays starting in ionized regions, and thus quantifies bubble sizes well even after ionized regions overlap.  To exclude un-resolved bubbles from the BSD, we do not count a cell as part of an ionized bubble unless it is $\geq 99\%$ ionized.  We caution that our simulations likely have too few resolved small bubbles - those with sizes $\sim$ a few $h^{-1}$Mpc - both due to our limited spatial resolution and our implementation of sub-resolved sources (see \S\ref{subsec:sources}).

Figure~\ref{fig:bubble_sinks} (top row) shows the IBSD at 20\%, 50\%, and 80\% (left to right) for our sinks models.  The IBSD confirms that the \maxC~has the smallest bubbles at all times, and that the other three models have similar bubble sizes.  The average bubble size is given at 20\%, 50\%, and 80\% ionized for each model in Table~\ref{tab:mean_bubble_radius}. The mean values are mainly intended to illustrate the relative differences between our models.  At $20\%$ and $50\%$ ionized the {\UrlFont Relaxed Limit} model has slightly larger bubbles, but at $80\%$ ionized is indistinguishable from the {\UrlFont Full Sinks} model.  The bubble sizes for the \unifC~model are slightly smaller than for the {\UrlFont Full Sinks} model, but are within $10-15\%$ at all times.  We see from the {\UrlFont Relaxed Limit} comparison that even assuming a fully pressure-smoothed IGM at all times is a reasonable approximation for morphology, especially late in reionization.  

\begin{table}
    \centering
    \begin{tabular}{|c|c|c|c|}
    \hline
       {\bf Mean Bubble Size [$h^{-1}$Mpc]} &  20\%  & 50\% & 80\%\\
       \hline
        {\UrlFont Full Sinks} & 1.94 & 7.30 & 30.08\\
        {\UrlFont Relaxed Limit} & 2.45 & 8.10 & 29.26\\
        \unifC~& 1.66 & 6.84 & 26.86\\
        \maxC~& 1.21 & 4.41 & 18.18\\
    \hline
    \end{tabular}
    \caption{Mean ionized bubble size at 20\%, 50\%, and 80\% ionized for each of the sinks models in this section.  }
    \label{tab:mean_bubble_radius}
\end{table}

\subsection{21 cm Power Spectrum}
\label{subsec:21cmpower}

The 21 cm power spectrum, which probes the \HI\ fluctuations in the IGM, is being targeted by PAPER~\citep{Parsons2010}, MWA~\citep{Tingay2013}, LOFAR~\citep{Yatawatta2013}, HERA~\citep{DeBoer2016,HERA2021a,HERA2021b}, and forthcoming experiments such as SKA~\citep{Koopmans2015}.  Ignoring redshift-space distortions and assuming the spin temperature of the 21 cm transition $T_S$ is much greater than the CMB temperature, we can write the 21 cm brightness temperature at position $\vec{x}$ as
\begin{equation}
    \label{eq:T21}
    T_{21}(\vec{x}) = \overline{T_{21}} x_{\rm HI}(\vec{x}) (1 + \delta(\vec{x}))
\end{equation}
where $\overline{T_{21}}$ is $T_{21}$ at mean density in neutral gas, which depends on redshift and cosmology only\footnote{Specifically, $\overline{T_{21}}^2 \propto 1 + z$.  Since our \maxC~model has a somewhat earlier re-ionization history, when comparing to that model we re-scale $\overline{T_{21}}$ to bring it to the same redshift as the other models.  Thus our comparisons reflect only differences sourced by $x_{\rm HI}$.  }, $x_{\rm HI}$ is the \HI\ fraction, and $1+\delta$ is the gas density.  The dimensionless 21 cm power spectrum is $\Delta_{21} \equiv k^3/2 \pi^2 P_{21}(k)$, where $P_{21}(k)$ is the power spectrum of $T_{21}$.  Since $\Delta_{21}$ depends on $x_{\rm HI}$, it is sensitive to the differences in morphology between our sinks models. 

Figure~\ref{fig:bubble_sinks} (bottom row) shows $\Delta_{21}$ vs. wavenumber $k$ for our sinks models at 20\%, 50\%, and 80\% ionized fractions (left to right).  In all cases we see familiar qualitative features.  Early on, $\Delta_{21}$ is steep in $k$ and its amplitude on large scales reaches a local minimum - a result of inside-out reionization~\citep{McQuinn2018,Giri2019b}.  
Later, $\Delta_{21}$ flattens out and its amplitude at $k \leq 0.2$ $h^{-1}$Mpc has increased by $1-2$ orders of magnitude by an ionized fraction of 80\%. (Note the different y axes on different panels.) This happens because the ionization field, which fluctuates on scales characteristic of the largest ionized bubbles ($10-30$ $h^{-1}$Mpc), takes over for the density field as the main driver of $\Delta_{21}$ at small $k$.  
Note that in this and in subsequent sections, we only show $\Delta_{21}$ for $k \leq 0.6$ $h$Mpc$^{-1}$, due to the caveat regarding the effects of sub-resolved halos discussed in \S\ref{subsec:sources}.  

The main effect of sinks is to reduce $\Delta_{21}$ on large scales ($k \leq 0.6$ hMpc$^{-1}$) by decreasing the sizes of large ionized bubbles.  At $20\%$ ionized, $\Delta_{21}$ at $k = 0.1 h\text{Mpc}^{-1}$ for the ({\UrlFont Relaxed Limit}, \maxC) model is ($1.75$, $0.59$) times its {\UrlFont Full Sinks} model value.  At $50\%$ ionized these numbers become ($1.16$, $0.43$), and at $80\%$ ionized, they are ($1.0$, $0.69$).  In all panels the {\UrlFont Full Sinks} and \unifC~models are always within a few percent of each other.  We see that the \maxC~model, which neglects the effects of pressure smoothing, under-estimates the large-scale $\Delta_{21}$ by $\approx 30-60\%$ relative to the {\UrlFont Full Sinks} case during much of reionization.  The {\UrlFont Relaxed Limit} over-estimates the power substantially only at $20\%$ ionized, and becomes an increasingly better approximation as reionization progresses.   

The \maxC~model illustrates that neglecting pressure smoothing can lead to a significant under-estimate of the large-scale power, owing to that model's smaller ionized bubbles.  On the other hand, assuming a fully relaxed IGM likely over-estimates the power early on, but becomes a reasonable approximation in the last half of reionization. Finally the similarity of the {\UrlFont Full Sinks} and \unifC~models suggests that $\Delta_{21}$ is unlikely to be sensitive to the details of how sinks are modeled, as long as the dynamics of the sinks can be accounted for in an average fashion via a uniform sub-grid clumping factor.   We caution, however, that all of these conclusions are sensitive to the properties of the sources, and we have employed only our fiducial source model so far.  As we will see in \S\ref{subsec:sourcesink}, the impact of sinks becomes larger (smaller) when fainter (brighter) sources dominate the photon budget.  

\subsection{Neutral Islands}
\label{subsec:neutralislands}

\begin{figure}
    \centering
    \includegraphics[scale=0.105]{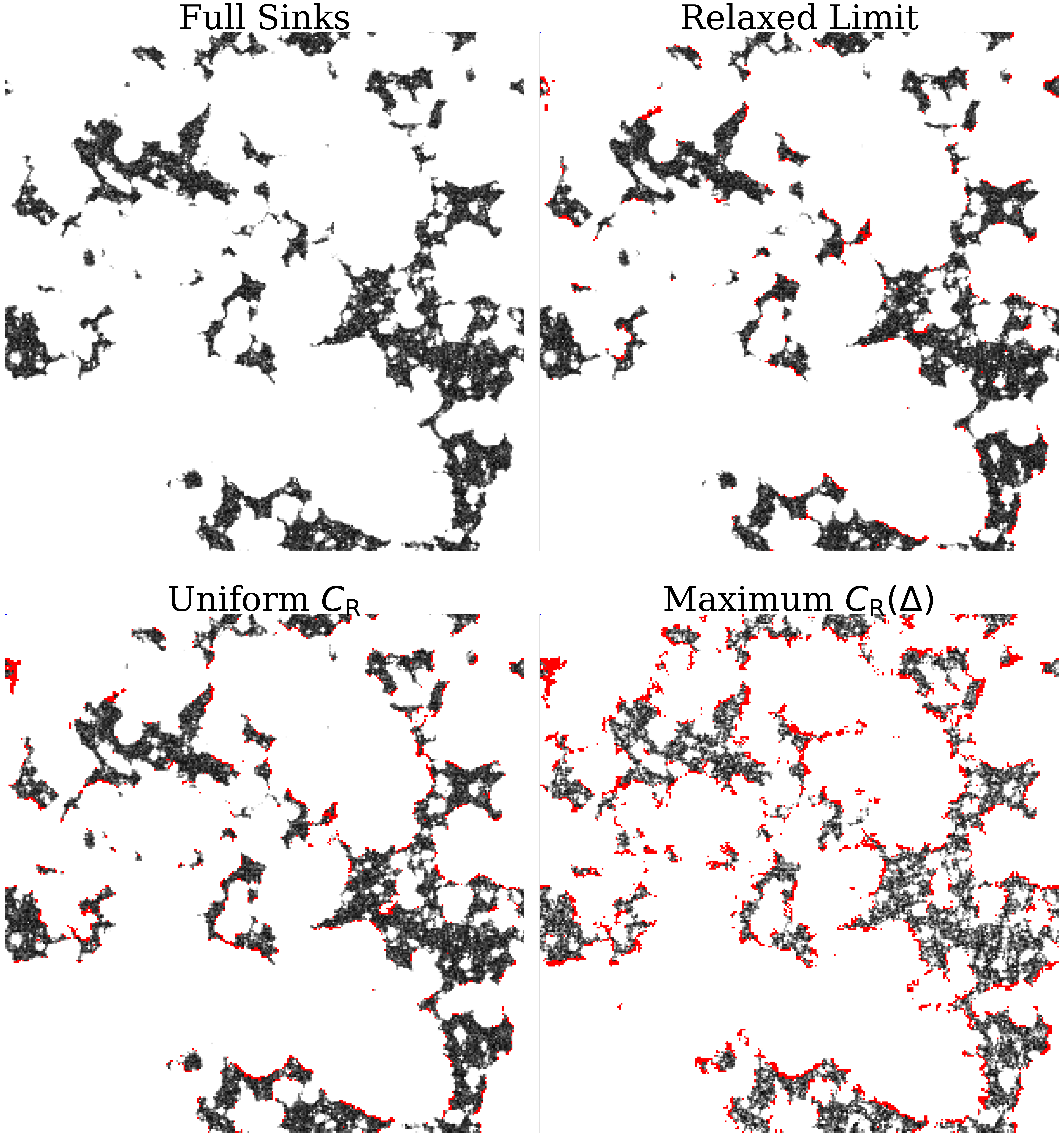}
    \caption{Visualization of neutral islands at 10\% volume neutral fraction for our sinks models, assuming our fiducial source scenario.  To aid comparison, the red shading highlights neutral regions that are ionized in the {\UrlFont Full Sinks} model.   The {\UrlFont Full Sinks}, {\UrlFont Relaxed Limit} and \unifC~models have visually similar island morphologies, while the \maxC~ case has more extended and more fragmented islands.  }
    \label{fig:netural_vis_sinks}
\end{figure}

So far our focus has been the morphology of ionized bubbles during the bulk of reionization. However, a lot of progress toward understanding reionization is being made with the growing number of $z>5$ QSO absorption spectra, which may be probing the final phases of reionization when the mostly ionized IGM was punctuated by islands of neutral gas.\footnote{These probes include Ly$\alpha/\beta$ forest statistics from QSO spectra~\citep{Fan2006,Becker2015,McGreer2015,Bosman2021,Zhu2021,Zhu2022}, the mean free path~\citep{Worseck2014,Becker2021,Bosman2021b}, and the LAE-forest connection~\citep{Becker2018, Meyer2020, Christenson2021, Ishimoto2022}.  }  Here we will briefly explore the morphology of these ``neutral islands''.  Neutral islands have been the focus of a number of recent studies~\citep[e.g.][]{Xu2014,Malloy2015,Xu2018,Giri2019,Wu2021b} owing to their importance for late-reionization observables.  

In Figure~\ref{fig:netural_vis_sinks}, we illustrate the distribution of neutral gas at 10\% volume neutral fraction using slices through our simulations.  The red shading in each panel corresponds to neutral regions that are ionized in the {\UrlFont Full Sinks} model, i.e. to highlight differences in the neutral island morphology with that model.  We see that the neutral islands in the {\UrlFont Relaxed Limit} and \unifC~models differ very little from the {\UrlFont Full Sinks} case, while there are substantial differences with the \maxC~model.  In that model the neutral structures are more extended -- as illustrated in red -- but also appear to be a lighter shading of gray. This lighter shading indicates that the neutral islands are more porous, i.e. they contain a larger number of small ionized bubbles inside of them.  

We quantify the morphology with the neutral island size distribution (NISD), defined analogously to the IBSD.  Late in reionization, the NISD is sensitive to the definition of a ``neutral'' cell, since most of the cells with neutral gas are partially ionized, especially in models with high opacity.  We define a cell to be part of an island if $x_{\rm HI} > 0.01$.  This choice is motivated by the fact that a sightline intersecting a partially neutral cell must pass within $1$ Mpc/h of an ionization front.  Gas this close to I-fronts typically has a low photo-ionization rate~\citep{Nasir2020} and/or is un-relaxed~\citep{Park2016,DAloisio2020} and is thus likely to be opaque to both LyC and Ly$\alpha$ photons.  

\begin{figure}
    \centering
    \includegraphics[scale=0.205]{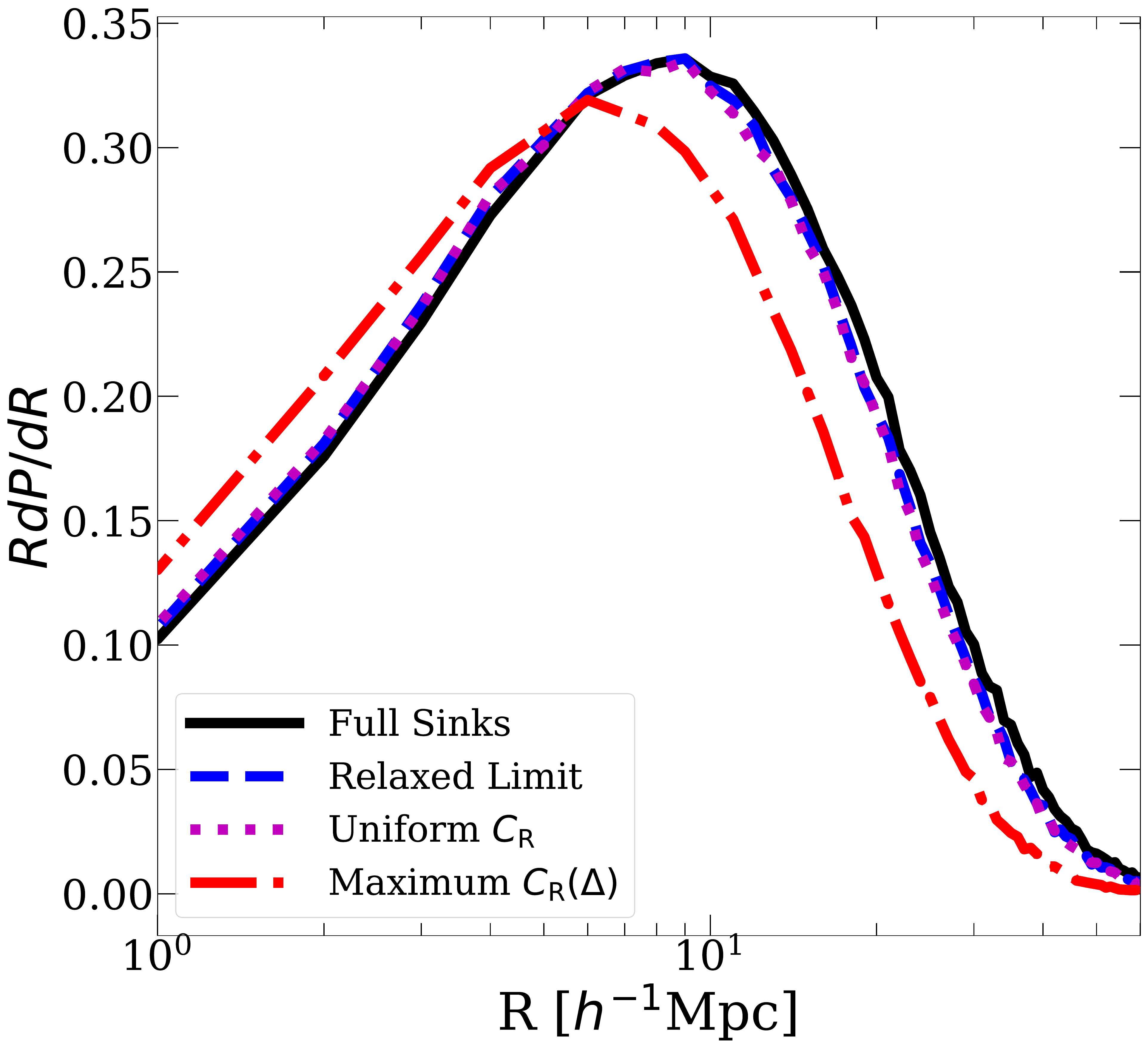}
    \caption{Neutral island size distribution defined such that any cell with $x_{\rm HI} > 0.01$ is part of an island.  We include all the sinks models in Figure~\ref{fig:bubble_sinks} and adopt our fiducial source model. The \maxC~model has smaller islands, while the NISDs for the other three models are all very similar.  }
    \label{fig:island_sinks}
\end{figure}

 Figure~\ref{fig:island_sinks} shows the NISD at 10\% volume neutral fraction for our sinks models (which occurs at $z \approx 5.5$ for all models except the \maxC~case, which is shown at $z \approx 6.0$).   The \maxC~model has the smallest islands while the other models are all very similar.  The average island sizes for the {\UrlFont Relaxed Limit}, {\UrlFont Full Sinks}, \unifC~and \maxC~models are $7.5$ $h^{-1}$Mpc,~$7.9$ $h^{-1}$Mpc,~$7.4$ $h^{-1}$Mpc,~and $5.8$ $h^{-1}$Mpc,~respectively.  In spite of the \maxC~model having more spatially extended neutral structures, the large abundance of small ionized bubbles within these structures break them up and shift the NISD towards smaller sizes.  We see that even the approximation of a fully pressure-smoothed IGM is likely acceptable for capturing the morphology of neutral islands.  On the other hand, ignoring pressure smoothing effects leads to a $\approx 20\%$ under-estimate of the mean island size in our fiducial source model.  

\section{Interplay Between Sources and Sinks}
\label{sec:sources}

\subsection{Source Models}
\label{subsec:imsource}

\begin{figure*}
    \centering
    \includegraphics[scale=0.15]{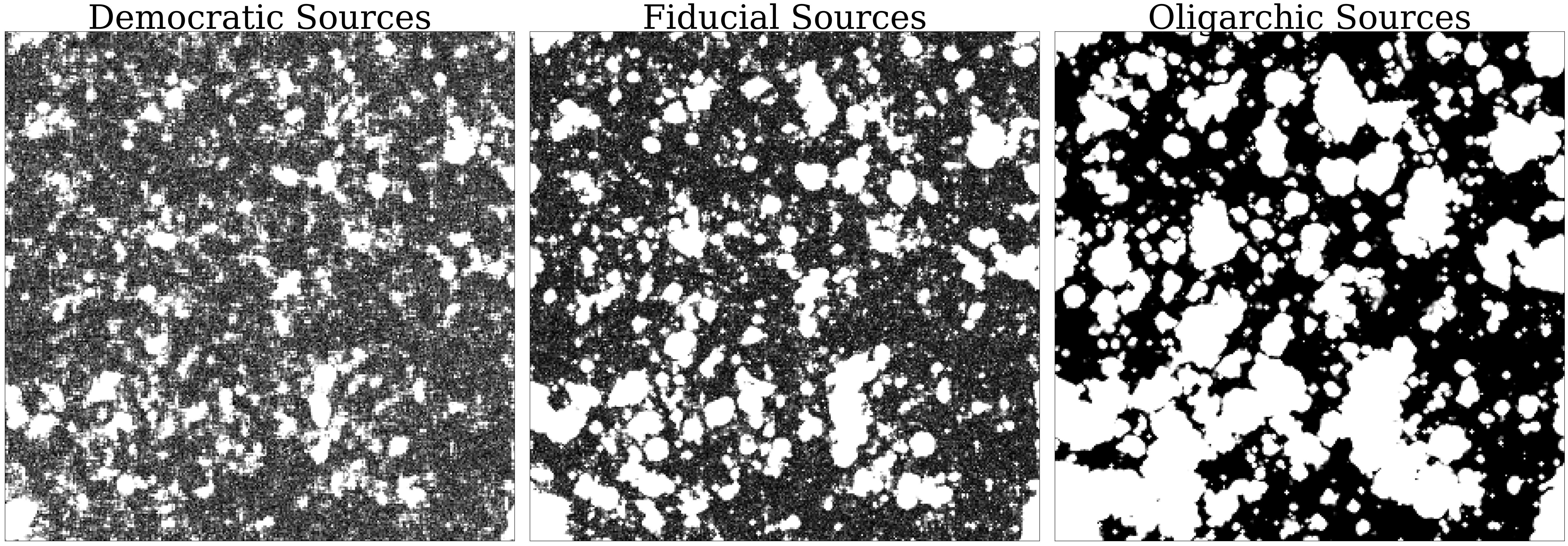}
    \caption{Ionization maps at 50\% volume ionized for the {\UrlFont Democratic Sources} (left), Fiducial (middle) and {\UrlFont Oligarchic Sources} (right) models, all assuming the {\UrlFont Full Sinks} model.  From left to right, the contribution to the photon budget from the brightest, most massive halos increases.  Reionization by more massive, highly biased sources leads to ionized bubbles being larger and fewer in number.  }
    \label{fig:ion_vis_sources1}
\end{figure*}

In this section, we will generalize our analysis to include different models for the sources.  Most previous studies of morphology have varied the source and sinks properties one at a time, while keeping the other fixed~\citep[e.g.][]{McQuinn2007,Shukla2016,Mao2019,Giri2019,Wu2021b,Chen2022}.  Our use of efficient RT simulations with sink dynamics included allows us to explore the relationship between the sources and sinks as it pertains to morphology.  We consider three models for the sources: 

\begin{itemize}

    \item {\UrlFont Democratic Sources}: This model differs from our fiducial model in that it assigns all halos the same ionizing emissivity independent of their luminosity, i.e. $\beta = 0$ (see \S\ref{subsec:sources}).  At $z = 6$, 50\% of the ionizing emissivity is produced by halos in the mass range $10^9 < M/[{\rm h}^{-1}M_{\odot}] < 1.8\times10^9$ ($-12.6 > M_{\rm UV} > -13.4$).  This model was introduced in~\citet{Cain2021} in an attempt to find a model that better recovers the short mean free path at $z = 6$ reported by \citet{Becker2021}.  This kind of picture would require a steep dependence of $f_{\rm esc}$ and/or the ionizing efficiency $\xi_{\rm ion}$ on luminosity, specifically, $f_{\rm esc} \xi_{\rm ion} \propto L_{\rm UV}^{-1}$ (corresponding to roughly $f_{\rm esc}\xi_{\rm ion} \propto M^{-1.4}$ over most of the mass range at $z = 6$).  The sources driving reionization in this model are almost entirely below current detection limits, in contrast to the {\UrlFont Oligarchic Sources} model described below.  
    
    \item {\UrlFont Fiducial Sources}: Our fiducial scenario with $M_{\min} = 10^9$ $h^{-1}M_{\odot}$ and with the emissivity of each halo proportional to its UV luminosity (i.e. $\beta = 1$).  At $z = 6$, halos with masses in the range $10^9 < M/[{\rm h}^{-1}M_{\odot}] < 1.8\times10^{10}$ ($-12.6 < M_{\rm UV} < -16.9$) contribute $50\%$ of the ionizing emissivity.  Of our three source models, this one is most similar to parameterizations commonly used in simulations, e.g. those that assume the emissivity to be proportional to halo mass~\citep{Mao2019,Keating2019,Keating2020,Bianco2021}.    
    
    \item {\UrlFont Oligarchic Sources}: In this model, bright and massive sources -- the ``oligarchs'' -- dominate reionization.  We adopt $M_{\min} = 2 \times 10^{10}$ $h^{-1}M_{\odot}$ with $\beta = 1$, corresponding to a limiting magnitude of $M^{\max}_{\rm UV}(z = 6) \approx -17$, roughly the limit of current observations at $5 < z < 10$~\citep{Finkelstein2016,Bouwens2021}. Thus it assumes that the sources responsible for reionization have, for the most part, already been observed.  This model is qualitatively similar to that proposed by~\citet{Naidu2020} (see also~\citet{Naidu2022,Mattee2022}).  
    It also serves to contrast starkly with the {\UrlFont Democratic Sources} model.   
\end{itemize}

To make some contact with previous works exploring how the source properties affect morphology, Figure~\ref{fig:ion_vis_sources1} shows ionization maps at 50\% volume ionized ($z \sim 7$) for our {\UrlFont Democratic Sources} (left), {\UrlFont Fiducial Sources} (middle) and {\UrlFont Oligarchic Sources} (right), all assuming the {\UrlFont Full Sinks} model.  The differences are clearly visible in the ionization fields; in the models driven by brighter sources, the ionized bubbles are larger and fewer in number.  This is because the most massive, rare sources produce a larger fraction of the photons in the {\UrlFont Fiducial} and {\UrlFont Oligarchic Sources} models.  This familiar result has been observed in many previous studies~\citep[e.g.][]{McQuinn2007, Giri2019,Kannan2022,Chen2022}. Now we turn our attention to the interplay between the sources and sinks.    

\subsection{Results}
\label{subsec:sourcesink}

\begin{figure*}
    \centering
    \includegraphics[scale=0.166]{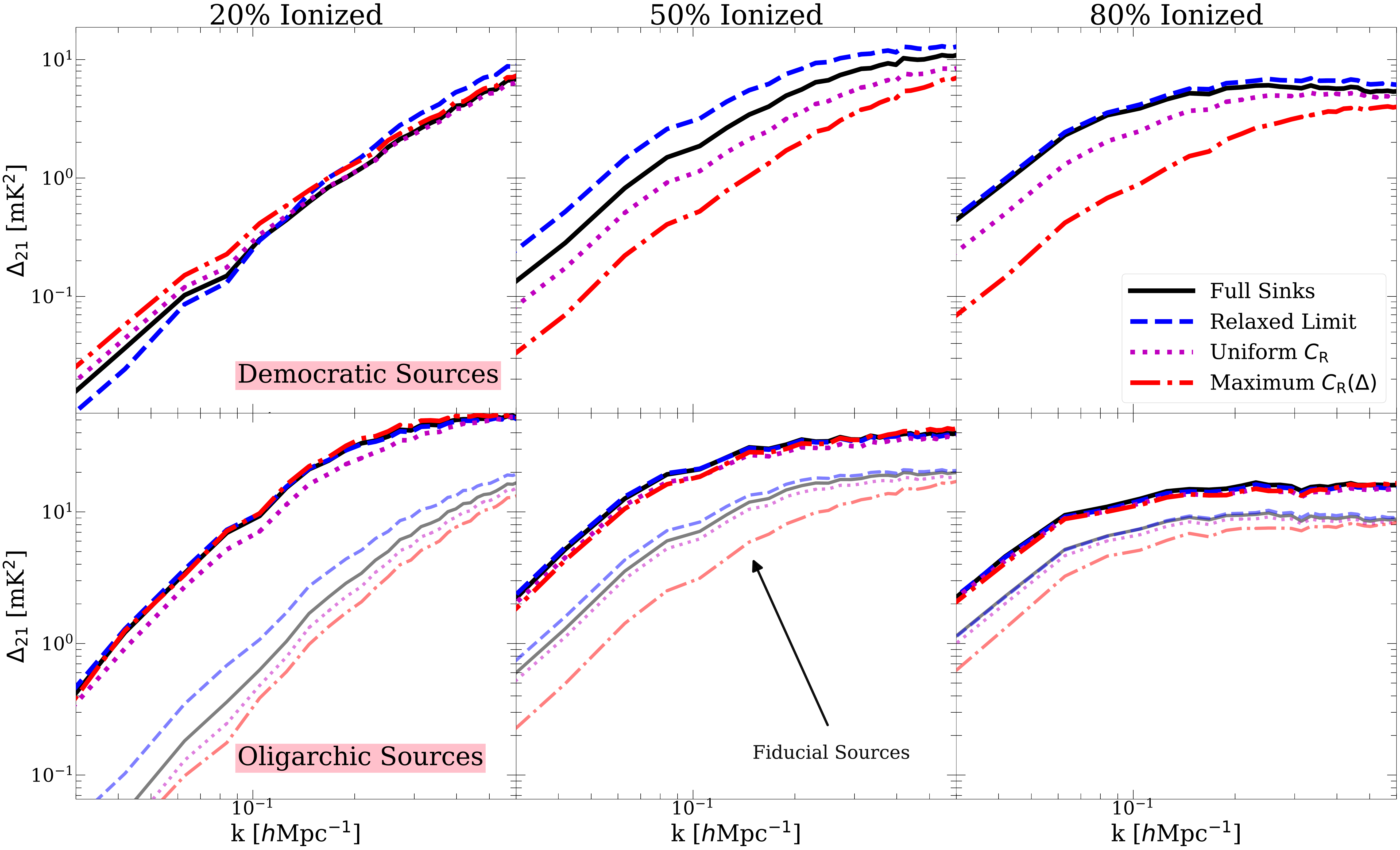}
    \caption{$\Delta_{21}(k)$ for all our sources and sinks models, illustrating the interplay between the sources and sinks of reionization.  
    The top and bottom rows show results for the {\UrlFont Democratic Sources} and {\UrlFont Oligarchic Sources} models, respectively.  The {\UrlFont Fiducial Sources} result is shown again by the thin curves in the bottom panel for comparison.  The magnitude of the sinks' effect on morphology clearly depends on the nature of the sources. In the {\UrlFont Democratic Sources} scenario (top row), the differences between the sinks models are significantly larger than in the {\UrlFont Oligarchic Sources} scenario.   Notably, in the former, the \unifC~model no longer matches so well the {\UrlFont Full Sinks} results, as it did for {\UrlFont Fiducial Sources} (compare to the thin curves in the bottom row). By contrast, in the {\UrlFont Oligarchic Sources} scenario, even the \maxC~model matches the others well, especially at 50\% and 80\% ionization.  We are led to conclude that in reionization scenarios driven by less (more) biased sources, the sinks become more (less) important for shaping morphology. }
    \label{fig:power_vs_ion_z_21cm_sources_sinks}
\end{figure*}

Figure~\ref{fig:power_vs_ion_z_21cm_sources_sinks} shows $\Delta_{21}(k)$ at 20\%, 50\%, and 80\% ionized, in the same format as the bottom panel of Figure~\ref{fig:bubble_sinks}, for all combinations of source and sinks models.  The top and bottom rows show results for the {\UrlFont Democratic Sources} and {\UrlFont Oligarchic Sources} models, while the results for the {\UrlFont Fiducial Sources} model (same as Figure~\ref{fig:bubble_sinks}) are shown by the thin lines in the bottom row.  Note that models sharing the same sinks prescription have similar reionization histories and the same emissivity histories as those shown in Fig.~\ref{fig:ion_history_sinks}.  In the {\UrlFont Democratic Sources} case, the differences between sinks models are smaller at $20\%$ ionized but somewhat larger at $50\%$ and $80\%$ ionized than in the {\UrlFont Fiducial Sources} case.  The suppression of $\Delta_{21}$ at $k = 0.1$ $h$Mpc$^{-1}$ in the latter half of reionization relative to the {\UrlFont Full Sinks} case has increased from $30-60\%$ for the {\UrlFont Fiducial Sources} case to $70-80\%$.  In addition, there are now $\approx 40\%$ differences between the {\UrlFont Relaxed Limit}, {\UrlFont Full Sinks}, and \unifC~models at $50\%$ ionized.  The \unifC~model is $\approx 35\%$ below the {\UrlFont Full Sinks} and {\UrlFont Relaxed Limit} models at $80\%$ ionized.  

It is interesting that for the {\UrlFont Democratic Sources} model (top row) the {\UrlFont Full Sinks} and \unifC~models have appreciably different $\Delta_{21}$.  In particular, the {\UrlFont Full Sinks} model has {\it more} large-scale power, which is indicative of larger ionized bubbles.  Recall from our discussion in \S\ref{subsec:hydrovis} that the {\UrlFont Full Sinks} model should be expected to favor the growth of larger bubbles more than the \unifC~case owing to lower (higher) clumping factors in over-dense (under-dense) regions in the former.  It seems that the these differences, which had little effect on morphology in our {\UrlFont Fiducial Sources} model, do become important in the limit that very faint, low-bias sources drive reionization.  We caution that this effect may be exaggerated due to our probable over-estimation of the impact of un-relaxed gas, discussed in \S\ref{subsec:caveats}.  However, it may also be a slight under-estimate due to the effects of using sub-resolution sources, as discussed in \S\ref{subsec:sources}.  In our tests using the {\UrlFont Democratic Sources} + \unifC~ combination, fixing the positions of the sources (see last paragraph of \S \ref{subsec:sources}) can reduce power at $k = 0.1$ $h$Mpc$^{-1}$ by up to $20\%$, while the {\UrlFont Full Sinks} model does not change appreciably.  This reduction in power was as large as a factor of $2$ in our tests using the {\UrlFont Democratic Sources} + \maxC~ combination.  We note that these differences would work in the direction of strengthening our conclusions in these scenarios, and that for the other source models we found differences of $10\%$ or less\footnote{Indeed, the {\UrlFont Oligarchic Sources} model does not use sub-resolution sources.  }.  

By contrast, in the {\UrlFont Oligarchic Sources} case, the differences are $15\%$ or less between the {\UrlFont Full Sinks}, {\UrlFont Relaxed Limit}, and \unifC~models in all of the panels.  More strikingly, at $50\%$ and $80\%$ ionized even the \maxC~model is very similar to the \unifC~case\footnote{In the {\UrlFont Oligarchic Sources} scenario, the earlier ionization history in the \maxC~model may obscure morphological differences that would be present if it had the same reionization history as the other sinks models.  This is because the bias of the sources evolves strongly with redshift in the {\UrlFont Oligarchic Sources} model due to its high $M_{\min}$.  To check this, we ran a {\UrlFont Relaxed Limit} simulation with an accelerated reionization history similar to the \maxC~one.  We found evidence for mild suppression (at most $20\%$ at $k = 0.1$ $h$Mpc$^{-1}$) at $50\%$ ionized, and no sign of suppression at $80\%$ ionized.  This is less than the effect seen in the {\UrlFont Fiducial Sources} case, confirming our statement in the text.  }.  The insensitivity of morphology to the sinks in the {\UrlFont Oligarchic Sources} model contrasts the much stronger dependence seen in the {\UrlFont Democratic Sources} model.  

Why is morphology sensitive to the sinks in models driven by fainter sources, but not in the {\UrlFont Oligarchic Sources} scenario?  In \S\ref{sec:results}, we saw that sinks limit the sizes of large ionized bubbles.  However, it is harder for them to do so in the {\UrlFont Oligarchic Sources} scenario for two reasons.  Nearly all the emissivity is concentrated in highly biased regions, strongly favoring the growth of the largest ionized bubbles.  Second, these bubbles grow fast enough to escape the over-densities in which they are born before recombinations begin having a significant impact.  This mitigates the ``disadvantage'' those bubbles have of inhabiting over-dense regions.  In these ways, sources in the {\UrlFont Oligarchic Sources} model ``win out'' over the sinks in terms of shaping morphology.  In the {\UrlFont Democratic Sources} model, by contrast, the sources are less biased than in {\UrlFont Fiducial Sources} and the sinks can more easily slow the growth of the largest bubbles.  In other words, the sinks are unable to tax the rich enough to affect morphology when the source bias is very high, and become more effective at taxing them when the source bias is reduced.  

This result has implications for forthcoming efforts to model reionization and interpret observations.  Most straightforwardly, it demonstrates that studying the sinks and sources one at a time can produce biased results.  For example, studying the sinks in a scenario with only highly biased sources would lead to the incorrect conclusion that they are unimportant for morphology.   
Another point is that very highly-biased source models may be relatively easy to rule out (or confirm) with forthcoming 21 cm observations from reionization.  For example, an upper limit of e.g. $\Delta_{21}(k = 0.1 \text{ hMpc}^{-1}) \leq 10$ mK$^2$ midway through reionization would strongly disfavor the {\UrlFont Oligarchic Sources} model (which has $\Delta_{21}(k = 0.1 \text{ hMpc}^{-1}) \approx 20$ mK$^{2}$ at $x_{\rm ion}^{\rm V} = 0.5$), since any physically reasonable sinks model would be unable to push $\Delta_{21}$ much lower than this\footnote{This statement presumes that at fixed ionized fraction, only the sources and sinks appreciably impact morphology.  Two other effects - redshift-space distortions~\citep{Ross2021} and spin temperature fluctuations~\citep{HERA2021b} may also impact the observed signal significantly.  However, both of these work to boost large-scale power, which would only strengthen our statement about upper limits.  }.  The tightest upper limit to date from HERA~\citep{HERA2021a} is $\Delta_{21} \leq 946$ mK$^{2}$ at $z \sim 7.9$ and $k = 0.19$ $h$Mpc$^{-1}$, less than 2 dex away from reaching the prediction of our {\UrlFont Oligarchic Sources} model. Other probes that are sensitive to the existence of large ionized regions, such as the visbility of LAEs at $z > 6$~\citep{Vanzella2011,Jung2020,Tilvi2020,Endsley2021}, may also be able to identify large bubbles like those predicted by the {\UrlFont Oligarchic Sources} model.  

\section{Conclusion} \label{sec:conc}

At present, there is no consensus on how much of an effect the sinks had in shaping reionization's morphology and, relatedly, how important they are for interpreting its observables.  We have attempted to address these questions using cosmological RT simulations of reionization.  Our simulations include the sub-grid model for the ionizing photon opacity developed by \citet{Cain2021}, which is based on high-resolution, fully coupled radiative hydrodynamics simulations of the IGM. The model improves over previous efforts in several key ways: it includes the effects of self-shielding and hydrodynamic response to photo-heating, keeping track of their dependencies on the LyC intensity, the timing of (local) reionization, and the environmental density.   Our main conclusions can be summarized as follows:

\begin{itemize}

    \item The sinks decrease the sizes of the largest ionized bubbles during reionization. We explored this effect in our detailed sub-grid model ({\UrlFont Full Sinks}), and in three other models representative of the ways that sinks have been implemented in previous studies: (1) A model that assumes a pressure-smoothed IGM ({\UrlFont Relaxed Limit}); (2) A simple clumping factor without dynamics or spatial in-homogeneity, tuned to have the same photon budget as our fiducial model (\unifC); (3) An in-homogeneous clumping model from \citet{Mao2019} that neglects pressure smoothing, thus representing a kind of upper limit on the effects of sinks (\maxC).  
    
    \item For our fiducial source model, which assumes the same escape fraction and ionizing efficiency for all sources, the {\UrlFont Full Sinks} model has up to $\sim 10-20\%$ smaller mean bubble sizes compared to the {\UrlFont Relaxed Limit} model in the first half of reionization. These differences mostly disappear in the second half.    
    
    \item By contrast, the \maxC~model underestimates bubble sizes by $\approx 40\%$ (compared to the {\UrlFont Full Sinks} model).  Ignoring the dynamical effects of pressure smoothing and photoevaporation can over-estimate significantly the sinks’ effects on morphology.  
    
    \item 
    We were able to reproduce a very similar morphology to our {\UrlFont Full Sinks} model using a uniform constant sub-grid clumping factor (the \unifC~model).  Hence, under typical assumptions about reionization's source population, with regards to morphology, it appears that the detailed dynamics and spatial in-homogeneity of the sinks can be adequately modeled in an average sense with a sub-grid clumping factor.  This is a useful result for scenarios where either (1) the ionizing photon budget is fixed by a model (as in this work) or by some empirical constraint, or (2) the budget is free to vary, as in a parameter space study.  To apply this result to a reionization simulation, one may simply re-scale the recombination rates at $T_{\rm ref}=10,000$ K by a uniform sub-grid clumping factor, $C_{\rm R}$, tuned to match the given total ionizing photon budget.  Note, however, that the {\UrlFont Full Sinks} and \unifC~models exhibit significant differences in $\Gamma_{\rm HI}$ (Fig. \ref{fig:ion_history_sinks}), which could render predictions for, e.g., the Ly$\alpha$ forest quite different.  As such, we emphasize that this conclusion should only be taken to apply to the structure of ionized and neutral regions, and not other physical properties of the ionized IGM, such as $\Gamma_{\rm HI}$ or the mean free path.

    \item {Differences in bubbles sizes between our models are manifest in the predicted power spectrum of the red-shifted 21cm background.  The \maxC~under-estimates the large-scale 21 cm power by $30-60\%$ throughout reionization compared to our {\UrlFont Full Sinks} model for our fiducial source prescription.  The {\UrlFont Relaxed Limit} model over-estimates power somewhat early in reionization, but becomes similar to both the {\UrlFont Full Sinks} and \unifC~cases in reionization's latter half.  }
    
    \item {The morphology of neutral islands near the end of reionization is very similar in all of the models except the \maxC~case, which produces smaller islands.  The islands in that model are $20\%$ too small on average, highlighting again the importance of including the effects of pressure smoothing.  }
    
    \item The strength of the sinks' effect on morphology is sensitive to the properties of the sources that drove reionization.  In a model where reionization was driven entirely by bright ($M_{\rm UV} < -17$), highly biased galaxies, the sinks suppress the 21 cm power at the $10-15\%$ level at a fixed ionized fraction throughout reionization, even in the \maxC~case.  By contrast, when faint ($M_{\rm UV} \sim -13$), low-bias  galaxies drove reionization, the large-scale 21 cm power can be suppressed by up to $80\%$, and the morphology in the {\UrlFont Full Sinks} and \unifC~models differ significantly.  This result highlights the need to study the effects of sinks and sources together instead of separately.  Moreover, the insensitivity of morphology to sinks in highly biased source models makes such models easier targets for forthcoming 21 cm experiments like HERA and SKA, and other probes sensitive to the presence of very large ionized bubbles. 
    
\end{itemize}

Our {\UrlFont Full Sinks} model can be improved on in several ways.  First, in future iterations we plan to address the caveats discussed in \S\ref{subsec:caveats}, namely the possible under-counting of rare, massive sinks and double-counting of absorptions in self-shielded systems.  These issues can be addressed with sub-grid simulations in larger volumes and by explicitly modeling the evolution of the residual \HI\ fraction in self-shielded systems.  A notable uncertainty in our results is that simulations upon which our sub-grid model is based do not include galaxy formation processes, which may affect significantly the structure and state of sinks near massive halos.  

Given the interplay between sources and sinks pointed out here, future studies should also move beyond simplistic source parameterizations.  Source models should ideally incorporate physically motivated prescriptions for effects such as feedback from reionization~\citep{Shapiro1994,Thoul1996,Gnedin2000,Hoeft2006,Finlator2011,Wu2019b,Ocvirk2021}, bursty star formation~\citep[][]{Weisz2011,Emami2019,Furlanetto2022}, galaxy formation histories~\citep[][]{Bullock2000,Somerville2015,Mirocha2021}, and for $f_{\rm esc}$~\citep{Kuhlen2012,Barrow2020,Maji2022,Marques-Chaves2022,Yeh2022}, all of which play important roles in setting the abundance and bias of the sources.  

\section*{Acknowledgements}

We thank Simeon Bird for his help running MP-Gadget, and Hy Trac for providing the SCORCH simulation results against which we calibrated our source models. A.D.’s group is supported by NASA 19-ATP19-0191, NSF AST-2045600, and JWST-AR-02608.001-A. M.M. also acknowledges  NASA 19-ATP19-0191 .  All computations
were made possible by NSF XSEDE allocation TG-PHY210041 and the NASA HEC Program through the NAS Division at Ames Research Center.

\section*{Data Availability}

The data underlying this article will be shared upon reasonable request to the corresponding author.



\bibliographystyle{mnras}
\bibliography{references} 



\appendix

\section{Numerical Convergence}
\label{app:conv}

\begin{figure}
    \centering
    \includegraphics[scale=0.26]{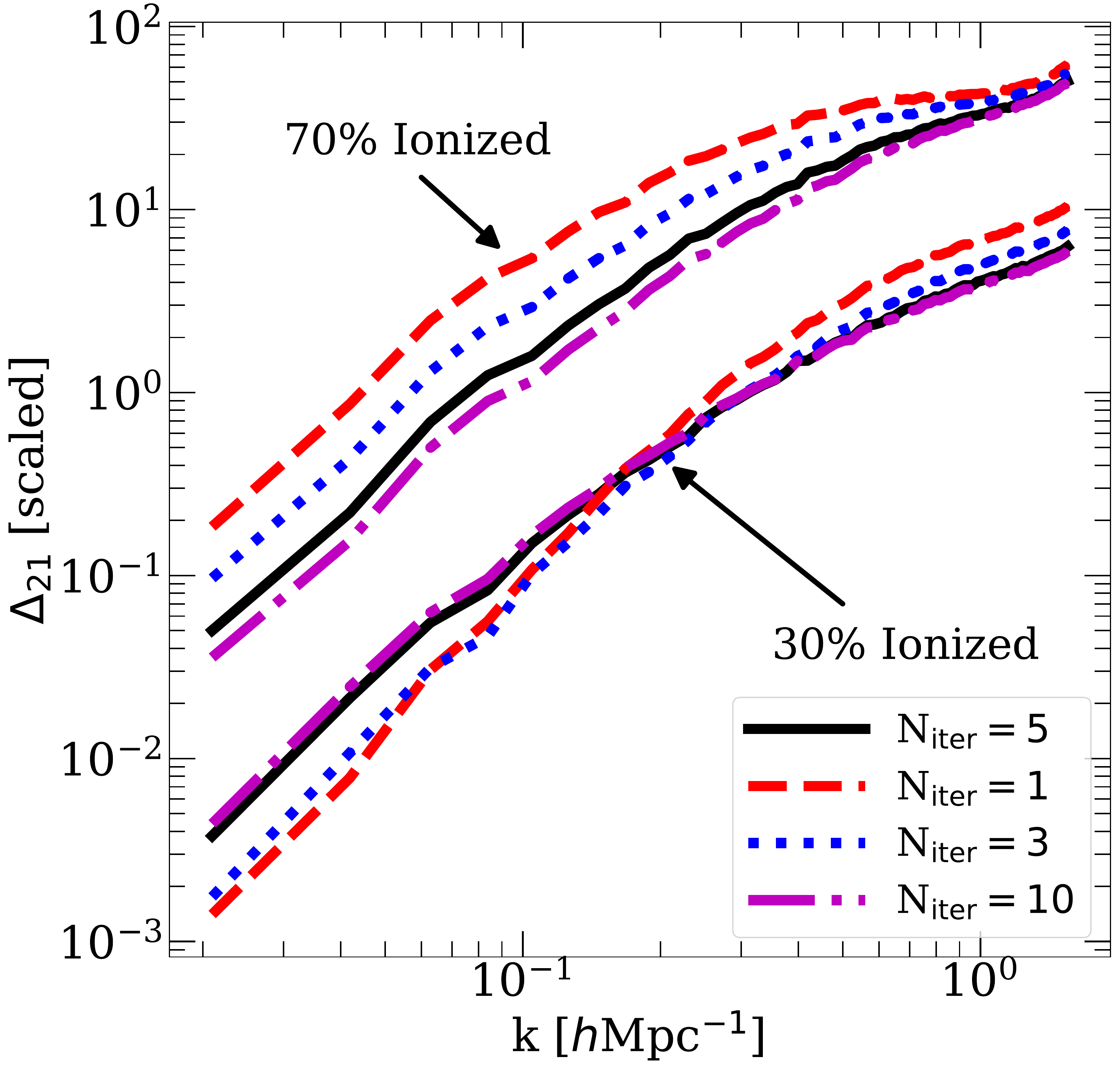}
    \caption{$\Delta_{21}(k)$ at 30\% and 70\% ionized (see annotations) for four values of $N_{\rm iter}$ (see legend).  This test uses the {\UrlFont Democratic Sources} and \maxC~models (the combination with the shortest $\overline{\lambda}$) in a coarse-grained version of the original simulation volume with $2$ $h^{-1}$Mpc cells.    Our fiducial value of $N_{\rm iter} = 5$ is within 10\% or the $N_{\rm iter} = 10$ case at $30\%$ ionized and within $35\%$ at $70\%$ ionized at all $k$.  This is relatively small compared to the differences seen in the top row of Figure~\ref{fig:power_vs_ion_z_21cm_sources_sinks}. 
    Moreover, all the other combinations of models that we checked displayed significantly better convergence.  }
    \label{fig:conv_tests}
\end{figure}

Here we describe some additional parameters in our code and demonstrate convergence of the ionization field in our simulations.  The first parameter is $N_{\rm iter}$, the number of times Eq.~\ref{eq:gammah1_subgrid} is iterated with the equation for $\overline{\lambda}$ (Eq.~\ref{eq:lambdamaster} or~\ref{eq:clumping_factor_model}) during each time step.  Our fiducial value is $N_{\rm iter} = 5$.  Our initial guess for $\Gamma_{\rm HI}$ assumes $\overline{\lambda} >> \Delta x_{\rm cell}$, where $\Delta x_{\rm cell}$ is the cell size, in which limit Eq.~\ref{eq:gammah1_subgrid} is independent of $\overline{\lambda}$.  Thus in general, convergence takes longest when $\overline{\lambda} \lesssim \Delta x_{\rm cell}$ - that is, in optically thick cells.   To test convergence of $N_{\rm iter}$, we ran simulations with $N_{\rm iter} = 1$, $3$, $5$, and $10$ on a coarse-grained ($N = 150^3$; $\Delta x_{\rm cell} = 2~h^{-1}$Mpc) version of our reionization volume using the {\UrlFont Democratic Sources} and \maxC~models.  This is the most extreme combination of source and sinks scenarios since it has the shortest $\overline{\lambda}$ on average.  In Fig.~\ref{fig:conv_tests} we show $\Delta_{21}(k)$ vs. wavenumber at 30\% and 70\% ionized for our tests, re-scaled so that the two sets of curves can be distinguished.  At $k = 0.1$ $h$Mpc$^{-1}$, the $N_{\rm iter} = 5$ and $10$ cases are $\approx 10\%$ apart at $30\%$ ionized and $35\%$ apart at $70\%$ ionized.  This is considerably less than the factor of several difference between the \maxC~and {\UrlFont Full Sinks} models the top row of Figure~\ref{fig:power_vs_ion_z_21cm_sources_sinks}.  We have checked convergence for different combinations of sinks and source models and found better convergence in all cases.  Moreover, this result is conservative because the condition $\overline{\lambda} \lesssim \Delta x_{\rm cell}$ is more likely to occur for $\Delta x_{\rm cell} = 2$ $h^{-1}$Mpc than for our fiducial $\Delta x_{\rm cell} = 1$ $h^{-1}$Mpc.     

Next we checked for convergence in the angular resolution of the radiation field.  This is adjustable in our code through two parameters that control how rays are merged.  The first, $l_{\rm hpx}$, is the order of the HealPix sphere onto which rays are binned when they are merged.   Our fiducial $l_{\rm hpx} = 0$ corresponds to keeping track of $12$ directions.  The other parameter is $N_{\rm ex}$ - the number of rays per cell that are ``exempt'' from being merged.  Before rays merged, they are sorted in order of their photon counts, and the top $N_{\rm ex} N^3$ rays are not considered for merging\footnote{We found that this procedure considerably reduces noise in the radiation field, particularly around the brightest sources.  }.  Using the same coarse-grained setup, we checked all combinations of $l_{\rm hpx} = 0$ and $1$ (which corresponds to tracking $48$ directions) and $N_{\rm ex} = 16$ (our fiducial choice) and $44$.  We found that $\Delta_{21}$ for these tests (not shown) to be indistinguishable for all combinations of these parameters on scales of interest, despite the amount of noise in the radiation field decreasing considerably for higher resolution runs. 

\section{Derivation of Eq. 1 (for $\Gamma_{\rm HI}$)}
\label{app:gamma_deriv}

Here we will derive Eq.~\ref{eq:gammah1_subgrid} for $\Gamma_{\rm HI}$.  Consider cell $i$ with ionized fraction $x_{\rm ion}^{i}$ and volume $V_{\rm cell}$.  If the I-front in cell $i$ is infinitely sharp and travels along one axis, then ray $j$ intersecting cell $i$ will travel a distance $x_{\rm ion}^{i} \Delta s^{ij}$ (recall $\Delta s^{ij}$ is the total path length of ray $j$ through cell $i$) before reaching neutral gas.  The number of photons absorbed over this distance is
\begin{equation}
    \label{eq:nabs}
    N_{\rm abs}^{i} = \sum_{j=1}^{N_{\rm rays}} N_{0}^{ij} \left(1 - \exp\left[\frac{-x_{\rm ion}^{i} \Delta s^{ij}}{\overline{\lambda}^{i}}\right]\right)
\end{equation}
where $N_{0}^{ij}$ is the number of photons in ray $j$ entering cell $i$ and $\overline{\lambda}$ is the mean free path in cell $i$ behind the I-front.  During a time step $\Delta t$, $\Gamma_{\rm HI}$ behind the I-front is
\begin{equation}
    \label{eq:gammai}
    \Gamma_{\rm HI}^{i} = \frac{\text{\# of photons absorbed per time}}{\text{\# of HI atoms in ionized gas}} = \frac{N_{\rm abs}^{i}/\Delta t}{n_{\rm HI}^{\Gamma} x_{\rm ion}^{i} V_{\rm cell}}
\end{equation}
where $x_{\rm ion}^{i} V_{\rm cell}$ is the ionized volume of cell $i$ and
\begin{equation}
    \label{eq:nh1_gamma}
    n_{\rm HI}^{\Gamma} \equiv \frac{\langle \Gamma_{\rm HI} n_{\rm HI}\rangle_{\rm V}}{\langle \Gamma_{\rm HI} \rangle_{\rm V}}
\end{equation}
is the $\Gamma_{\rm HI}$-weighted HI number density (the V sub-script denotes a volume average).  Eq.~\ref{eq:lambda_freq} relates the numerator of Eq.~\ref{eq:nh1_gamma} to our definition for $\overline{\lambda}$ for the small-volume simulations (derived in the next section).  Combining Eqs.~\ref{eq:lambda_freq}, \ref{eq:nh1_gamma}, and \ref{eq:nabs} yields
\begin{equation}
    \label{eq:gammai_2}
    \Gamma_{\rm HI}^{i} = \sum_{j=1}^{N_{\rm rays}} \frac{ N_{0}^{ij} \left(1 - \exp\left[-x_{\rm ion}^{i} \Delta s^{ij}/\overline{\lambda}^{i}\right]\right)}{(\overline{\lambda}^{-1} F_{\gamma}/\langle \Gamma_{\rm HI} \rangle_{\rm V}) x_{\rm ion}^{i} V_{\rm cell} \Delta t}
\end{equation}
where $F_{\gamma} \equiv \Gamma_{\rm HI}^0 \overline{\sigma}_{\rm HI}^{-1}$ is the ionizing flux at the source planes in the small-volume simulations and $\Gamma_{\rm HI}^0$ is the photo-ionization rate at the source planes. 
Because the domain size (32 $h^{-1}$kpc) is much less than $\overline{\lambda}$ in all our small-volume simulations, $F_{\gamma}$ usually attenuates very little over the domain width except around self-shielded systems, which (typically) occupy a small fraction of the volume.  Thus, $\langle \Gamma_{\rm HI} \rangle_{\rm V} \approx \Gamma_{\rm HI}^0$, which gives
\begin{equation}
    \label{eq:gammai_final}
    \Gamma_{\rm HI}^{i} \approx \sum_{j=1}^{N_{\rm rays}} \frac{ N_{0}^{ij} \left(1 - \exp\left[-x_{\rm ion}^{i} \Delta s^{ij}/\overline{\lambda}^{i}\right]\right)}{(\overline{\lambda}\overline{\sigma}_{\rm HI})^{-1} x_{\rm ion}^{i} V_{\rm cell} \Delta t}
\end{equation}
which is equivalent to Eq.~\ref{eq:gammah1_subgrid}.

\begin{figure}
    \centering
    \includegraphics[scale=0.175]{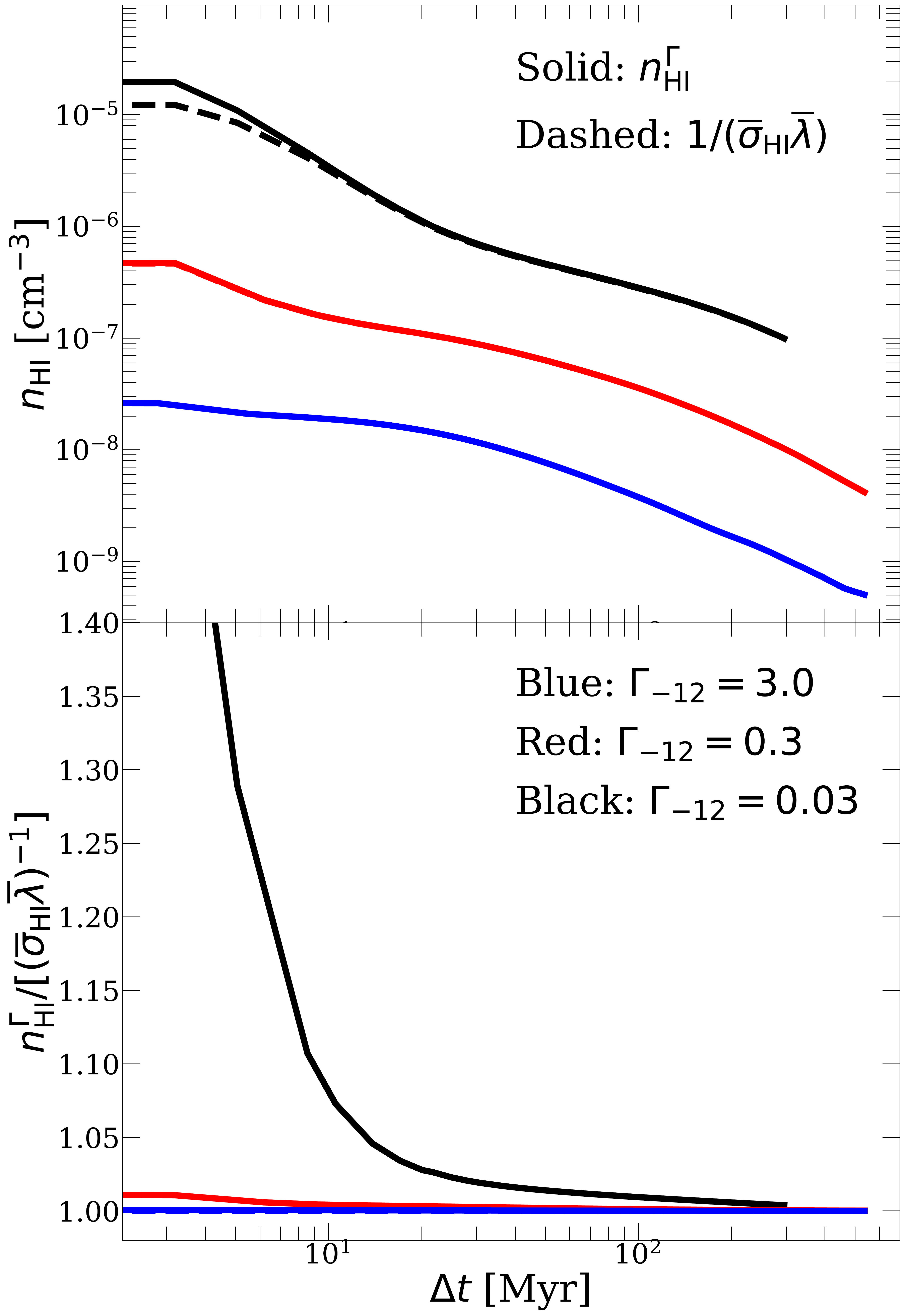}
    \caption{Test of the relation $n_{\rm HI}^{\Gamma} = (\overline{\lambda}\overline{\sigma}_{\rm HI})^{-1}$, as required by Eq.~\ref{eq:gammah1_subgrid}.  {\bf Top}: $n_{\rm HI}^{\Gamma}$ (solid) and $(\overline{\lambda}\overline{\sigma}_{\rm HI})^{-1}$ (dashed) vs. time since ionization for small-volume simulations with $\Gamma_{-12} = 3.0$ (blue), $0.3$ (red), and $0.03$ (black), assuming $z_{\rm re} = 8$ and $\delta/\sigma = 0$.  {\bf Bottom}: the ratio between these two quantities for each case.  For $\Gamma_{-12} = 3.0$ and $0.3$ the equality holds within a few percent at all times, but for $0.03$ agreement to within 10\% is not reached until $\Delta t \approx 10$ Myr.  This works in the direction of under-estimating the absorption rate in recently ionized cells with low $\Gamma_{\rm HI}$ in our reionization simulations.  }
    \label{fig:nh1gamma}
\end{figure}

Note that Eq.~\ref{eq:gammai} and Eq.~\ref{eq:gammai_final} together imply that $n_{\rm HI}^{\Gamma} \approx (\overline{\lambda}\overline{\sigma}_{\rm HI})^{-1}$ should be true in our small-volume simulations. Figure~\ref{fig:nh1gamma} tests this equality for simulations with $\Gamma_{-12} = 3.0$ (blue curves), $0.3$ (red) and $0.03$ (black) for $z_{\rm re} = 8$ and $\delta/\sigma = 0$ (mean density).  The top panel plots both quantities vs. time since ionization, while the bottom panel shows their ratio.  In the simulations with $\Gamma_{-12} = 3.0$ and $0.3$ the equality holds to within a few percent even during the first few Myr when self-shielding is most important.  However in the $0.03$ case, they do not agree to within 10\% until $\approx 10$ Myr after ionization.  In that case, $n_{\rm HI}^{\Gamma} > (\overline{\lambda}\overline{\sigma}_{\rm HI})^{-1}$, Eq.~\ref{eq:gammah1_subgrid} under-estimates the number of absorptions in ionized gas because it over-estimates $\Gamma_{\rm HI}$, and therefore the converged value of $\overline{\lambda}$ (Eq.~\ref{eq:lambdamaster}).  This works in the direction of making the opacity too low in recently ionized gas with low $\Gamma_{\rm HI}$ in our reionization simulations.  However, the double-counting issue described in \S\ref{subsec:caveats} likely still renders the total opacity in these regions an over-estimate.  The test and photon budget comparison described in that section includes the effect discussed here, so our statements there should still hold.  

\section{Derivation of Eq.2 (for $\overline{\lambda}$)}
\label{app:lambda_deriv}

In this section we derive our estimator for the frequency-averaged mean free path in our small-volume simulations, $\overline{\lambda}$ (Eq.~\ref{eq:lambda_freq}).  Let $I_\nu$ be the specific intensity at the source planes.  The ionizing flux along one direction of our box is,
\begin{equation}
F_{\gamma} =  \int^{4 \nu_{\rm HI}}_{\nu_{\rm HI}} \frac{I_\nu}{h_p \nu}~d\nu,
\end{equation}
where $h_p$ is Planck's constant and $h_p \nu_{\rm HI}$ is the ionization potential of hydrogen. Assuming the radiation streams along the $x_1$ direction, the photoionization rate at location $\mathbf{x}=(x_1,x_2,x_3)$ along a ray is
\begin{equation}
\Gamma_{\rm HI}(\mathbf{x}) = \int^{4 \nu_{\rm HI}}_{\nu_{\rm HI}} \frac{\sigma_\nu }{h \nu} I_\nu e^{-\int^{x_1}_0 dx' n_{\rm HI}(x',x_2,x_3) \sigma_\nu} ~d\nu,
\end{equation}
where $n_{\rm HI}(\mathbf{x})$ is the proper hydrogen number density and $\sigma_\nu$ is its photoionization cross section. We can write
\begin{equation*}
n_{\rm HI}(\mathbf{x})~\Gamma_{\rm HI}(\mathbf{x}) =  \int^{4 \nu_{\rm HI}}_{\nu_{\rm HI}} d \nu \frac{I_\nu}{h \nu} ~n_{\rm HI}(\mathbf{x})\sigma_{\nu}~ e^{-\int^{x_1}_0 dx' n_{\rm HI}(x',x_2,x_3) \sigma_\nu} =
\end{equation*}
\begin{equation}
-\int^{4 \nu_{\rm HI}}_{\nu_{\rm HI}} d \nu \frac{I_\nu}{h \nu} \frac{\partial }{\partial x_1} e^{-\int^{x_1}_0 dx' n_{\rm HI}(x',x_2,x_3) \sigma_\nu}
\label{eq:c3}
\end{equation}  
Integrating over the domain volume $V_d = L_d^3$, we obtain
\begin{equation*}
V_d \langle n_{\rm HI} \Gamma_{\rm HI} \rangle_{V_d} = 
\end{equation*}
\begin{equation}
\int^{4 \nu_{\rm HI}}_{\nu_{\rm HI}} d \nu \frac{I_\nu}{h \nu} \int_0^{L_d} dx_2 dx_3~\left( 1 -  e^{-\int^{L_d}_0 dx' n_{\rm HI}(x',x_2,x_3) \sigma_\nu} \right)
\label{eq:c4}
\end{equation}
where $\langle \ldots \rangle_{V_d}$ denotes an average over the domain volume.  We define the effective optical depth through
\begin{equation*}
e^{-\tau_{\mathrm{eff}}} \equiv \langle e^{-\int^{L_d}_0 dx' n_{\rm HI}(x',x_2,x_3) \sigma_\nu} \rangle_{A_d} = 
\end{equation*}
\begin{equation}
A_d^{-1} \int_0^{L_d} dx_2 dx_3~ e^{-\int^{L_d}_0 dx'n_{\rm HI}(x',x_2,x_3) \sigma_\nu}
\label{eq:c5}
\end{equation} 
where $A_d=L_d^2$ and $\langle \ldots \rangle_{A_d}$ denotes an average over the transverse plane.  Plugging this into equation \ref{eq:c4} yields
\begin{equation}
L_d \langle n_{\rm HI} \Gamma_{\rm HI} \rangle_{V_d} =  \int^{4 \nu_{\rm HI}}_{\nu_{\rm HI}} d \nu \frac{I_\nu}{h \nu} (1-e^{-\tau_{\mathrm{eff}}}).
\label{eq:c6}
\end{equation}
The mean free path is defined to be $\overline{\lambda} = L_d/\tau_{\mathrm{eff}}$. Assuming that $\lambda \gg L_d$ (recall that $L_d=32 h^{-1}$ kpc), we can expand the exponential in equation \ref{eq:c6} to first order, yielding
\begin{equation}
\label{eq:c7}
 \overline{\lambda}^{-1} \equiv \langle \lambda^{-1} \rangle_\nu = \frac{\langle n_{\rm HI} \Gamma_{\rm HI} \rangle_{V_d} }{F_{\gamma}},
\end{equation}  
where we have used that $\langle \lambda^{-1} \rangle_\nu = (1/F_{\gamma})\int^{4 \nu_{\rm HI}}_{\nu_{\rm HI}} d \nu \frac{I_\nu}{h \nu} \lambda^{-1} $.  The RHS of Eq.~\ref{eq:c7} is the volume-averaged absorption rate divided by the incident flux, and is equivalent to the volume averaged absorption coefficient.  Note that Eq.~\ref{eq:c7} counts {\it all} absorptions within ionized regions, not just those balanced by recombinations.  

\section{Test of Eq. 3 (to account for evolving $\Gamma_{\rm HI}$)}
\label{app:evolving_gamma}

\begin{figure}
    \centering
    \includegraphics[scale=0.265]{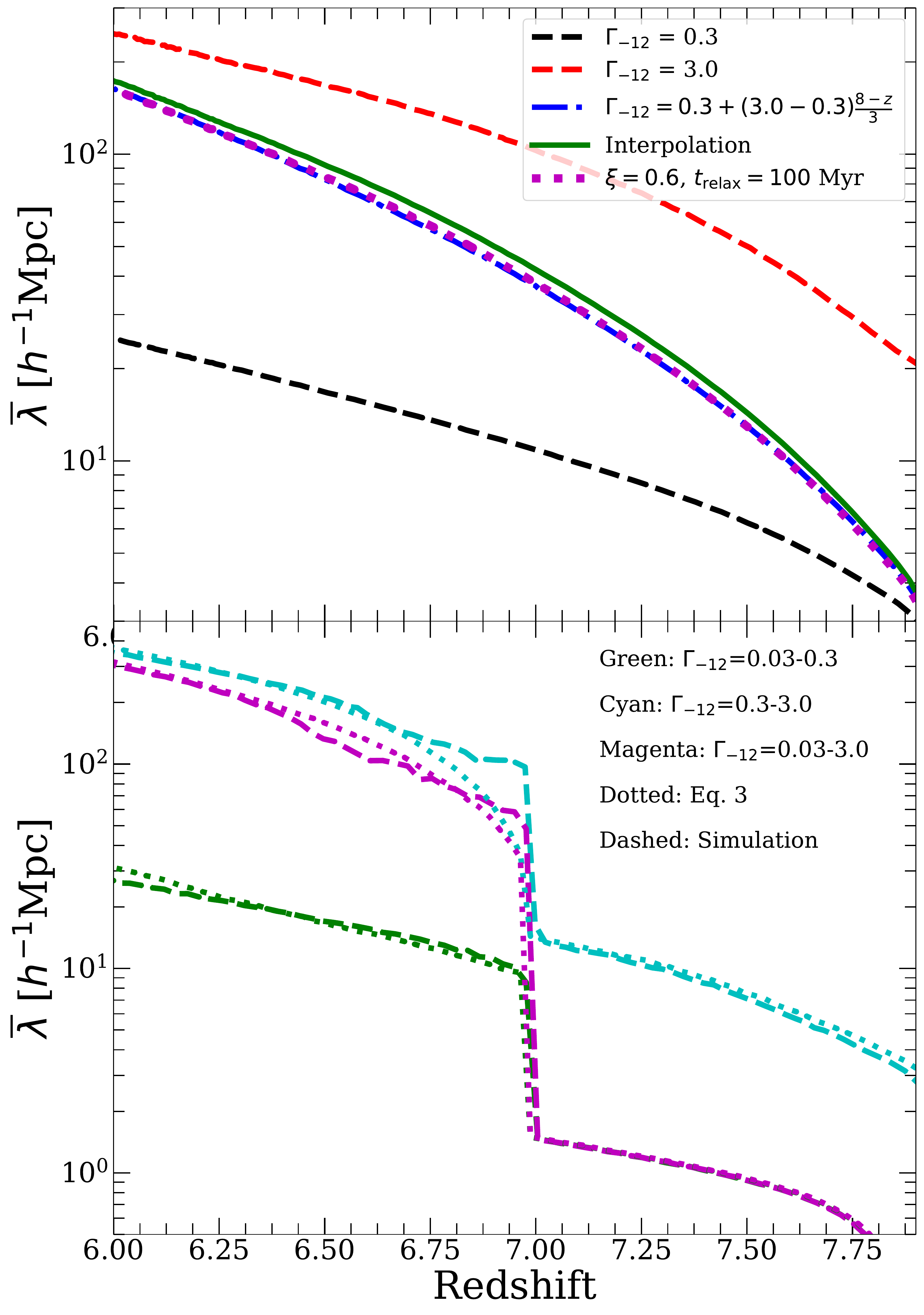}
    \caption{Tests of Eq.~\ref{eq:lambdamaster} using small-volume simulations with evolving $\Gamma_{\rm HI}$.  {\bf Top}: Test with $\Gamma_{\rm HI}(z) = 0.3 + (3.0 - 0.3)\frac{8 - z}{3}$ (blue dot-dashed) alongside a direct interpolation between simulations with constant $\Gamma_{\rm HI}$ (green solid) and the result of evaluating Eq.~\ref{eq:lambdamaster} with $\xi = 0.6$ and $t_{\rm relax} = 100$ Myr (magenta dotted).  The interpolation over-estimates $\overline{\lambda}$ by 10-15\% while Eq.~\ref{eq:lambdamaster} produces agreement to within a few percent.  {\bf Bottom}: Tests in smaller ($0.256$ $h^{-1}$Mpc) volumes in which we impulsively increased $\Gamma_{\rm HI}$ by $1-2$ orders of magnitude at $z = 7$.  The dashed curves are the simulation results and the dotted curves are Eq.~\ref{eq:lambdamaster}.  The model agrees reasonably well even in these extreme cases, although the values of $\xi$ and $t_{\rm relax}$ vary between fits (and from our fiducial values).  } 
    \label{fig:evolving_gamma}
\end{figure}

In this section we will show how Eq.~\ref{eq:lambdamaster} accounts for the sensitivity of $\overline{\lambda}$ to the history of $\Gamma_{\rm HI}$ in our RT cells.  Figure~\ref{fig:evolving_gamma} shows $\overline{\lambda}$ for several tests of Eq.~\ref{eq:lambdamaster} in small-volume simulations with evolving $\Gamma_{\rm HI}$.  In the top panel, we show the mean free path for a fiducial box size/resolution simulation with $\Gamma_{-12}(z) = 0.3 + (3.0 - 0.3)\frac{8 - z}{3}$ (dashed blue curve) alongside two approximations based on constant-$\Gamma_{\rm HI}$ simulations.  The solid green curve is a direct power law interpolation between $\Gamma_{-12} = 3.0$ (red dashed curve) and $0.3$ (black dashed curve) simulations.  The magenta dotted curve is the result of evaluating Eq.~\ref{eq:lambdamaster} with $\xi = 0.6$ and $t_{\rm relax} = 100$ Myr (close to our fiducial values for these parameters).  We see that the direct interpolation over-estimates $\overline{\lambda}$ by 10-15\%, while Eq.~\ref{eq:lambdamaster} agrees with the evolving $\Gamma_{\rm HI}$ simulation to within a few percent. 

\begin{table}
    \centering
    \begin{tabular}{|c|c|c|c|}
    \hline
       {\bf Starting $\Gamma_{-12}$} &  {\bf Ending $\Gamma_{-12}$}  & ${\bm \xi}$ & ${\bm t_{\rm relax}}$ [Myr]\\
       \hline
        0.03 & 0.3 & 0.8 & 700\\
        0.3 & 3.0 & 0.33 & 100\\
        0.03 & 3.0 & 0.67 & 300\\
    \hline
    \end{tabular}
    \caption{Best-fit parameters for our ``impulsive-$\Gamma_{\rm HI}$'' tests of Eq.~\ref{eq:lambdamaster}, shown in Figure~\ref{fig:evolving_gamma}.  }
    \label{tab:evolving_gamma_parameters}
\end{table}

We also ran several tests (in smaller boxes) in which we increased $\Gamma_{\rm HI}$ impulsively by $1-2$ orders of magnitude midway through the simulation.  These tests represent a ``maximum stress test'' of Eq.~\ref{eq:lambdamaster}, since in reality $\Gamma_{\rm HI}$ will evolve more gradually.  The bottom panel of Fig.~\ref{fig:evolving_gamma} shows the result of three tests, with $\Gamma_{\rm HI}$ impulsively jumping between the values quoted in the legend at $z = 7$.  The dashed lines show the simulation results and the dotted lines the result of Eq.~\ref{eq:lambdamaster} (evaluated using a suite of similar simulations with constant $\Gamma_{-12}$).  Though the values of $\xi$ and $t_{\rm relax}$ that gave these fits, given in Table~\ref{tab:evolving_gamma_parameters}, are somewhat different from each other (and our fiducial model), the goodness of the fits demonstrates the ability of Eq.~\ref{eq:lambdamaster} to capture $\overline{\lambda}$ in a variety of environments accurately.  The variation may be due in part to the smaller box sizes of these tests and the fact that $\xi$ and $t_{\rm relax}$ are partially degenerate, but we also do expect that $\xi$ and $t_{\rm relax}$ should in general depend on $\Gamma_{\rm HI}$ (and, in principle, over-density and $z_{\rm reion}$).  Future work will be required to address the environmental dependence of $\xi$ and $t_{\rm relax}$ in more detail.  


\bsp	
\label{lastpage}
\end{document}